\documentclass[twocolumn, trackchanges, twocolappendix]{aastex7}

\usepackage{amsmath}
\usepackage{float}
\usepackage[normalem]{ulem}

\usepackage{ctable}

\begin{document}

\title{The Radio--X-ray Correlation of High-Redshift AGN: A Numerical Study of Inverse-Compton Scattering of the CMB Photons in Relativistic Jets}

\author[orcid=0009-0006-9732-4248]{Aditya Sharma}
\affiliation{Department of Astronomy, Astrophysics and Space Engineering, Indian Institute of Technology Indore, Khandwa Road, Simrol, Indore 453552, India}
\email[show]{phd2301121004@iiti.ac.in; adityas17013@gmail.com}  

\author[orcid=0000-0001-5424-0059]{Bhargav Vaidya} 
\affiliation{Department of Astronomy, Astrophysics and Space Engineering, Indian Institute of Technology Indore, Khandwa Road, Simrol, Indore 453552, India}
\email{bvaidya@iiti.ac.in}

\author[orcid=0000-0003-4747-4484]{Silvia Belladitta} 
\affiliation{Max-Planck-Institut für Astronomie, Königstuhl 17, 69117 Heidelberg, Germany}\
\affiliation{INAF – Osservatorio di Astrofisica e Scienza dello Spazio, Via Gobetti 93/3, I-40129, Bologna, Italy }
\email{belladitta@mpia.de}

\author[orcid=0000-0002-3528-7625]{Christian Fendt}
\affiliation{Max-Planck-Institut für Astronomie, Königstuhl 17, 69117 Heidelberg, Germany}
\email{fendt@mpia.de}

\author[orcid=0000-0001-5470-305X]{Dharam V. Lal}
\affiliation{National Centre for Radio Astrophysics - Tata Institute of Fundamental Research Post Box 3, Ganeshkhind P.O., Pune 411007, India}
\email{dharam@ncra.tifr.res.in}

\author[orcid=0000-0002-2931-7824]{Eduardo Bañados}
\affiliation{Max-Planck-Institut für Astronomie, Königstuhl 17, 69117 Heidelberg, Germany}
\email{banados@mpia.de}

\author[orcid=0000-0003-1922-9406]{Biman B. Nath}
\affiliation{Department of Physical Sciences, Indian Institute of Science Education and Research Mohali, Knowledge City, Sector 81, SAS Nagar, Punjab 140306, India}
\email{nath.biman@gmail.com}

\author[orcid=0009-0000-6497-3336]{Harshita Bhuyan}
\affiliation{Department of Astronomy, Astrophysics and Space Engineering, Indian Institute of Technology Indore, Khandwa Road, Simrol, Indore 453552, India}
\affiliation{Max-Planck-Institut für Astronomie, Königstuhl 17, 69117 Heidelberg, Germany}
\email{phd2201221002@iiti.ac.in}

\begin{abstract}

Relativistic jets from active galactic nuclei are expected to exhibit strong redshift evolution in their radiative output due to the increasing energy density of the cosmic microwave background (CMB). 
We investigate the role of inverse Compton (IC) scattering of CMB photons in regulating the radio and X-ray emission from large-scale jets using three-dimensional relativistic magnetohydrodynamic 
simulations coupled with a hybrid Eulerian–Lagrangian particle framework.
By keeping the jet dynamics and ambient medium properties fixed across redshifts, we are able to isolate the impact of the cosmological evolution of the CMB on the 
jet radiation.
From our simulations, we construct synthetic spectral energy distributions and intensity maps considering synchrotron and IC/CMB losses along with particle acceleration from shocks. 
We are able to reproduce the weak redshift dependence of radio luminosity and the strong enhancement of X-ray emission toward high redshift that is observed in radio-loud quasars. 
At high redshift, the X-ray luminosity follows the expected $(1+z)^4$ scaling, confirming IC/CMB as the dominant mechanism driving the X-ray enhancement. 
The resulting X-ray-to-radio flux ratio increases systematically with redshift and is consistent with observational constraints.
Finally, we show that slower jets exhibit a stronger redshift evolution of the X-ray enhancement than faster jets, highlighting the critical role of jet propagation length scales and particle energy evolution.
The simulations also naturally reproduce the steepening of the radio spectral index with redshift - the $\alpha$–$z$ relation - thus providing a unified framework that allows
to interpret the multiwavelength properties of high-redshift radio sources.
\end{abstract}

\keywords{\uat{Active galactic nuclei}{16} --- \uat{Radio Jets}{1347} --- \uat{High Energy astrophysics}{739} --- \uat{Relativistic Jets}{1390} --- \uat{X-ray quasars}{1821}}

\section{Introduction}
Active galactic nuclei (AGN) are extremely luminous sources found in the centers of galaxies, powered by the accretion of matter onto a supermassive black hole (SMBH). As material spirals inward through the accretion disk, it releases enormous amounts of energy, not just as radiation \citep{Shakura1973, 2014ARA&A..52..529Y}, but also in the form of powerful, relativistic jets \citep{Blandford1982, 2014ARA&A..52..589H, 2019ApJ...882....2V, 2019ARA&A..57..467B, 2021MNRAS.505.3596D}.

The family of AGNs have a particularly luminous subclass known as quasars \citep{Singh2013, 2013BASI...41....1K}, which have been observed up to redshifts $z \sim 7.5$ \citep{Banados2018,
2020ApJ...897L..14Y, 2018Natur.553..473B, 2020A&A...635L...7B, 2023ARA&A..61..373F, Wang2021}. 
Quasars host rapidly growing SMBHs and therefore serve as essential probes of black hole formation in the early universe \citep{Inayoshi2020, Johnson2016}. 
Such superluminous sources provide key insights into the chemical composition and interstellar medium of their host galaxies \citep{Dietrich_2003, Decarli2023} 
and play a crucial role in understanding galaxy evolution at high redshift. 

Although only 10 - 15 \% of quasars exhibit powerful, large-scale relativistic jets \citep{Padovani2017}, recent deep radio surveys suggest that jet activity in galaxy centers may be far more common, with compact or low-power jets potentially 
present in a substantial fraction of the quasar population \citep{2024A&A...691A.191C}. 
These jets interact with the ambient medium resulting in the feedback processes which triggers and quenches the star formation rates \citep{2012ARA&A..50..455F, Wagner2016, 2025Galax..13..102M}. 
The dominant non-thermal emission from jets spans the electromagnetic spectrum from radio to $\gamma$-rays.
The radio-optical emission of AGN jets is well explained as synchrotron radiation from relativistic electrons in the jet flow interacting with the jet's magnetic field. 

The origin of X-ray emission from jets, however, remains debated. 
The two main processes that may produce X-ray emission from AGN jets are 
(1) synchrotron emission from a secondary population of ultra-relativistic electrons with Lorentz factors ($\gamma$) in the range of $10^8$ \citep{Harris2006} and (2) 
inverse Compton scattering of cosmic microwave background (CMB) photons (IC/CMB) \citep{Tavecchio2000, Celotti2001}. 
Additional mechanisms such as synchrotron self-Compton \citep{Bloom1996} and inverse Compton scattering of photons from the accretion disk, the broad-line region, or the dusty torus \citep{Dermer1993, Sikora1994, Baejowski2003} 
may also contribute.

The discovery of kiloparsec scales X-ray jets with \textit{Chandra} initially favored the IC/CMB model \citep{Tavecchio2000}, which would require
bulk Lorentz factors of $\Gamma \simeq$ 10-20. 
However, \textit{Fermi} $\gamma$-ray observations challenged this interpretation at low redshift ($z<1$), as the strong $\gamma$-ray emission predicted by the IC/CMB model was not detected. 
This discrepancy motivated synchrotron-based explanations involving a secondary population of very high-energy electrons ($\gamma \simeq 10^8$-$10^9$) \citep{Meyer2015, Georganopoulos2016}. 
Nevertheless, IC/CMB remains a plausible mechanism at high redshift, where the CMB energy density increases as $(1+z)^4$, causing electrons to lose energy more rapidly and making it difficult for them to reach or maintain the extremely 
high energies required for synchrotron emission.

Several studies \citep{Schwartz2019, HodgesKluck2021, Ighina2022, Wu2013, Zhu2018, Zuo2024} report enhanced X-ray emission in high-redshift quasar jets relative to low-redshift systems, 
commonly interpreted as evidence for a significant contribution from IC/CMB  process. 
After all, the origin of X-ray emission from AGN jets still remains debated, and the relative importance of synchrotron emission and IC/CMB is still uncertain.
IC/CMB is expected to become efficient only beyond certain redshifts and jet length scales depending upon the bulk Lorentz factor and magnetic field energy 
density of the jet \citep{Schwartz2019, 2024Univ...10..227C}.

Beyond its role in shaping X-ray emission, the increasing CMB energy density with redshift is also expected to influence the radio properties of jets. 
A long-standing observational result in powerful extragalactic radio sources is that objects at higher redshift tend to exhibit systematically steeper radio spectra, i.e., the well-known $\alpha$–$z$ relation. 
This trend is particularly well established in high-redshift radio galaxies, where spectral steepening with increasing redshift has been consistently observed \citep{DeBreuck2000, Klamer2006}. 
Indeed, the ultrasteep-spectrum selection technique has often been used as an efficient method for identifying high-redshift radio galaxies and as a tracer of intermediate to high redshift ($z \geq 1$) systems \citep{2001MNRAS.326.1563J, 2014A&A...569A..52S, 2014MNRAS.443.2590S}. 
In contrast, no statistically significant $\alpha$–$z$ correlation has yet been established for high-redshift radio quasars \citep{2021MNRAS.508.2798S}. 
Several mechanisms have been proposed to explain the spectral steepening observed in the radio sources, including enhanced IC/CMB cooling, which preferentially depletes high-energy electrons at high redshift, and environmental effects, where denser ambient gas may intrinsically steepen the synchrotron spectrum \citep{Ghisellini2014, Athreya1998, Klamer2006}. 
However, despite extensive observational and theoretical efforts, the physical origin of the $\alpha$–$z$ relation remains uncertain, particularly in the quasar population, and continues to motivate further investigation.

From a theoretical perspective, modeling the coupled evolution of jet dynamics, particle acceleration, and radiative cooling is essential for interpreting emission signatures. 
This approach also enables the identification of the physical origin and spatial distribution of radiation within relativistic jets. 
Several studies have demonstrated the importance of jet dynamics in shaping shock structures, particle acceleration, and resulting emission signatures 
from pc \citep{Dubey2023, 2024ApJ...976..144D, 2024A&A...691A..14K, 2025A&A...699A.296S, 2026A&A...705A..74C, 2026MNRAS.546ag322S} 
to kpc \citep{Mukherjee2020, 2022A&A...667A.138K, Upreti2024, 2026arXiv260110787J, 2025PASA...42..136J, 2025A&A...703A.214G, 2026MNRAS.546ag131E} scale AGN jets. 
These studies highlight that continuous particle re-acceleration through multiple shocks is crucial for sustaining extended emission, 
and establish a framework that self-consistently links relativistic magnetohydrodynamic (RMHD) jet evolution with nonthermal radiation.

In this work, we extend this framework to investigate the evolution of relativistic jets at high redshift, with particular emphasis on kpc-scale emission and its dependence on cosmic epoch. We perform three-dimensional RMHD simulations using the \texttt{PLUTO} code \citep{Mignone2007}, coupled with a Lagrangian particle module \citep{Vaidya2018, Mukherjee2021} to model the transport and evolution of nonthermal electrons. 
Since the electron population extends to ultra-relativistic energies, IC scattering of the CMB photons is no longer well described by the Thomson approximation and requires a proper treatment of Klein–Nishina (KN) effects.
However, the existing implementation of the particle module does not include a complete treatment of IC/CMB emission and neglects KN effects, 
which are essential for modeling high-energy emission, particularly at high redshift. 
This paper extends the particle module to incorporate a full IC/CMB emissivity calculation, including relativistic transformations and KN suppression. 
This development represents a key numerical advancement of the present study. 
Essentially, this approach enables us to study the coupled evolution of radio and X-ray emission and to assess how the dominant X-ray emission mechanism 
changes as the jet evolves with both time and redshift.

The paper is organized as follows. 
In Section~\ref{sec:2}, we describe the numerical setup and simulation framework that governs the relativistic jet dynamics. 
Section~\ref{sec:3} presents the extensions made to the particle module to account for redshift-dependent effects, together with the development of a new inverse-Compton module capable of 
modeling emission from arbitrary isotropic blackbody photon fields. 
The updated particle framework is validated through direct comparison with one-zone calculations, such as those implemented in \texttt{AGNpy} \citep{Nigro2022}. 
In Section~\ref{sec:4}, we present the main results of the study, focusing on the dynamical evolution of jets in three-dimensional RMHD simulations and the associated synthetic spectral energy distributions across a range of 
redshifts.
Finally, Section~\ref{sec:5} discusses the implications of our results, with particular emphasis on the coupled evolution of radio and X-ray emission and its role in driving the observed X-ray enhancement in high-redshift radio-loud quasars.

Throughout this paper, we assume a flat $\Lambda$CDM cosmology with $H_0 = 67.74\ \mathrm{km\ s^{-1}\ Mpc^{-1}}$, $\Omega_m = 0.3$, and $\Omega_\Lambda = 0.69$ 
\citep{2016A&A...594A..13P}. 
At a redshift of $z = 5$, an angular scale of $1''$ corresponds to a projected physical distance of 6.43~kpc.

\section{Numerical and Theoretical Formulation} \label{sec:2}
We employ the \texttt{PLUTO} code \citep{Mignone2007} to perform three-dimensional simulations of a relativistic, rotating, and magnetized AGN jet on kpc scales.
The dynamical evolution of the plasma is governed by the RMHD equations, which the code solves in their conservative form,
\begin{equation}
\frac{\partial}{\partial t}
\left[
\begin{array}{c}
D \\
\mathbf{m} \\
E_t \\
\mathbf{B}
\end{array}
\right]
+ \nabla \cdot 
\left(
\left[
\begin{array}{c}
D \mathbf{v} \\
w_t \gamma^2 \mathbf{v} \mathbf{v} - \mathbf{b} \mathbf{b} + p_t \mathbf{I} \\
\mathbf{m} \\
\mathbf{v} \mathbf{B} - \mathbf{B} \mathbf{v}
\end{array}
\right]^T
\right)
= 0,
\end{equation}
where $D = \gamma \rho$ is the lab-frame mass density, with $\rho$ and $\gamma$ denoting the rest-mass density and the Lorentz factor respectively. 
The three-velocity is denoted by $\mathbf{v}$, and the covariant magnetic field is defined as 
$b^\mu = [b^0, \mathbf{b}] = [\gamma \mathbf{v} \cdot \mathbf{B}, \, \mathbf{B}/\gamma + \gamma (\mathbf{v} \cdot \mathbf{B}) \mathbf{v}]$. 
The total enthalpy is $w_t = \rho h + \frac{|\mathbf{B}|^2}{\gamma^2} + (\mathbf{v} \cdot \mathbf{B})^2$, where $h$ is the specific enthalpy.
The momentum density is expressed as $\mathbf{m} = w_t \gamma^2 \mathbf{v} - b^0 \mathbf{b}$, 
and the total pressure is defined by $p_t = p_g + \frac{1}{2} \frac{|\mathbf{B}|^2}{\gamma^2} + \frac{1}{2}(\mathbf{v} \cdot \mathbf{B})^2$, 
where $p_g$ is the thermal gas pressure. 
The total energy density in the lab frame is given by $E_t = w_t \gamma^2 - (b^0)^2 - p_t$, and $\mathbf{I}$ represents the identity tensor.

We solve the RMHD equations using the relativistic HLLD Riemann solver \citep{mignone2009} with piecewise parabolic spatial reconstruction \citep{Colella1984, Mignone2005} and a second-order Runge-Kutta time integrator. 
To preserve the solenoidal constraint on the magnetic field, $\nabla \cdot \mathbf{B} = 0$, we apply the hyperbolic divergence cleaning method proposed by \citet{Dedner2002}. 
For thermodynamics, we adopt the Taub-Matthews (TM) equation of state \citep{Mignone2007a}, which provides a consistent relation for relativistic flows and smoothly interpolates 
between the $\gamma=5/3$ (cold) and $\gamma=4/3$ (hot) limits. 
In this framework, the specific enthalpy is given by 
\begin{equation} \label{enthalpy}
  h = \tfrac{5}{2}\Theta + \sqrt{\tfrac{9}{4}\Theta^2+1},
\end{equation}
where $\Theta = p/\rho$ is the dimensionless temperature.

The simulations are carried out in a Cartesian domain $x = [-L_x,L_x]$, $y = [-L_y,L_y]$, $z = [0,L_z]$, with $L_x = L_y = 30\ l_0$ and $L_z = 160\ l_0$, 
corresponding to a total box size of $60\ l_0 \times 60\ l_0 \times 160\ l_0$. 
The equidistant grid size applied in our simulations is $360 \times 360 \times 960$.

In the \texttt{PLUTO} code, all variables are normalized using three fundamental scale parameters: the unit length $l_0$, the unit velocity $v_0$, and the unit density $\rho_0$. All other physical quantities are derived from these scales, including the unit time $t_0 = l_0/v_0$, magnetic field $B_0 = v_0\sqrt{4\pi\rho_0}$, pressure $p_0 = \rho_0 v_0^2$, and temperature $T_0 = \mu m_\mu v_0^2 / 2k_B$. We define the unit length as the jet radius, $l_0 = 0.1$~kpc, set the unit velocity to the speed of light, $v_0 = c$, and choose the unit density as $\rho_0$ = $1.661 \times 10^{-24} \ \rm{g \, cm^{-3}}$.

\subsection{Initial and Boundary Condition}\label{Initial and Boundary Condition}
We aim to simulate jets in high-redshift systems. 
We inject a fully relativistic, rotating, and magnetized jet injected through a cylindrical nozzle centered at the origin $(0,0,0)$, 
with radius $r_j = l_0$ and height $h_j = l_0$ in a relatively dense ambient environment. 
The ambient medium is assumed to be homogeneous and static, with density $\rho_0 = 1.661 \times 10^{-24}\ \mathrm{g\,cm^{-3}}$ and 
temperature $T = 10^7\ \mathrm{K}$ \citep{Mukherjee2020}. 
The jet-to-ambient density ratio is set to $\eta = 10^{-4}$, indicating that the jet is significantly lighter than the surrounding medium.

\begin{figure*}
    \centering
    \includegraphics[width=0.8\linewidth]{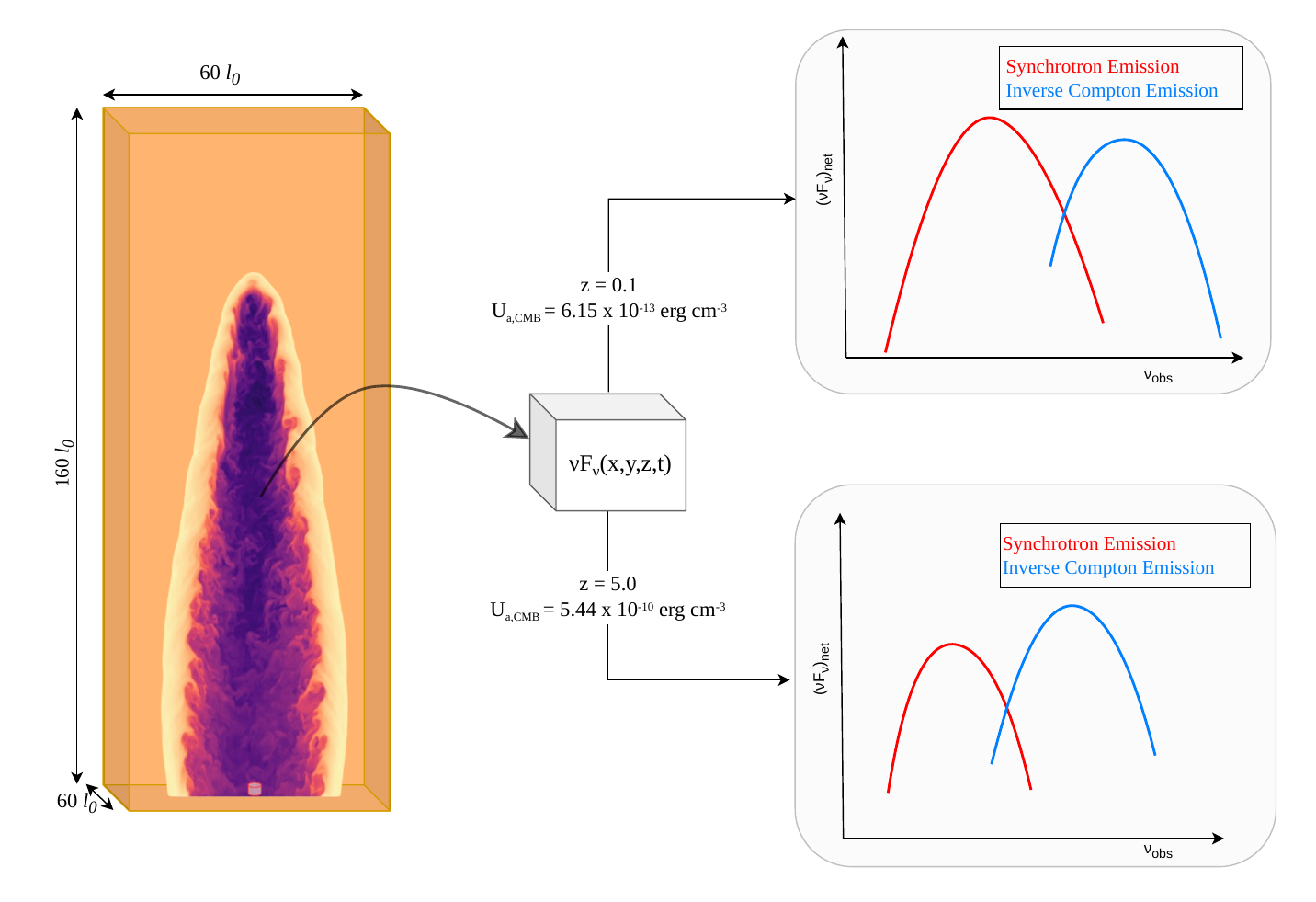}
    \caption{Visualization of the simulation setup and the central idea of this work. A relativistic, magnetized jet is injected into a uniform ambient medium (left panels) and evolves within a 3D domain of size $60\,l_0 \times 60\,l_0 \times 160\,l_0$ (center). 
    Using identical jet dynamics, we compute the expected emission at different redshifts, where the increasing CMB energy density enhances inverse-Compton emission relative to synchrotron emission. 
    The resulting change in the spectral energy distribution (right panels) illustrates how high-redshift jets may appear increasingly IC-dominated.}
    \label{fig:Geometry}
\end{figure*}

The nozzle parameters are chosen such that the jet initially maintains a radial equilibrium configuration, following the model of \citet{Bodo2019}, who adopted the steady-state, axisymmetric solution of the ideal RMHD equations. 
The choice of jet radius ensures that the radius of the jet inlet is resolved by at least six computational cells. 
The resolution applied here is less than that applied in e.g. \citet{Dubey2023}, however such a high resolutions is not feasible for the kpc-scale simulations we treat.
We thus follow \citet{Mukherjee2020} for our choice of resolution. 
A detailed description of the governing equations and the resulting radial profiles of the jet variables is provided in Appendix~\ref{Appendix 1}.

The ambient gas pressure is set to $P_a = 1.38 \times 10^{-9} \mathrm{dyne\,cm^{-2}}$. The jet is injected in pressure equilibrium with the ambient medium, and thus the nozzle pressure is initialized to match $P_a$. 
The ambient magnetic field, $B_a$, is uniform and oriented along the $\hat{z}$ direction, with magnitude $B_a = 5.476 \times 10^{-6} \, \mathrm{G}$. 
This choice matches the vertical component of the magnetic field inside the jet nozzle and ensures that the initial magnetic field configuration is divergence-free.

\subsection{Parameter Runs}
The focus of this work is to explore how the radio and X-ray emission from relativistic jets evolve with redshift, and to examine whether the rising CMB energy density plays a key role in driving these changes. In this regard, we adopt a simplified approach in which the jet dynamics are treated as independent of cosmic time. That is, for a given set of injection parameters, the jet evolves identically at all redshifts. The only quantity that varies with redshift is the CMB photon energy density, which scales as $(1+z)^4$ and directly influences radiative cooling processes such as synchrotron and inverse-Compton losses (see Section~\ref{sec:4}).

Under this framework, the injected jet is characterized by five key parameters specified along the central axis, that is the axial Lorentz factor $\Gamma_{c,j}$, 
the magnetic pitch  $P_{c,j} = r B_{z,j}/B_{\phi, j}$, 
the ratio of matter energy density to magnetic energy density $M_a^2 = \rho \Gamma_{c,j}^2 / {\langle B_j^2\rangle}$,
the strength of the centrifugal force relative to the magnetic pressure gradient $\alpha_j$, 
and the jet's angular velocity $\Omega_j$.
The parameters $P_{c,j} = 0.01$, $M_a^2=1000$, and $\alpha_j=1$ are fixed for all simulations.

Using these parameters, we perform seven simulations divided into two categories,
(a)~Rg5, with $\Gamma_{c,j} = 5$, consisting of four redshift-dependent runs,
and (b)~Rg2, with $\Gamma_{c,j} = 2$, consisting of three runs with varying redshift. 
The complete set of input parameters is listed in Table~\ref{tab:initial_param}.

In addition to jet injection parameters, the average jet power (excluding the rest-mass contribution) is also listed in the table and calculated as
\begin{multline}
    P_{0,j} = \pi r_j^2 v_{z,j} \Bigg[ 
        \Gamma_{z,j} (\Gamma_{z,j} - 1)\,\rho_j c^2 
        + \big((h_j - 1)\Gamma_z^2 - \Theta_j\big)\rho_j \\
        + \frac{|\mathbf{B}|^2}{2} 
        + \frac{|\mathbf{v}|^2 |\mathbf{B}|^2 - (\mathbf{v} \cdot \mathbf{B})^2}{2}
    \Bigg]
\end{multline}
\citep{Mignone2007a, English2016, Mukherjee2020, Upreti2024}.

\begin{table}[t]
\centering
\caption{
Input parameters for the jet injection nozzle (in code units).  
Columns list: (1) the run identifier;
(2) the redshift ($z$);
(3) the axial magnetic field component ($B_{zc,j}$);  
(4) the angular velocity at the jet axis ($\Omega_c$);  
(5) the average injected jet power ($\overline{P_{0,j}}$);  
(6) the electron number density ($N_e$); and
(7) the final simulation time ($t_{\rm final}$) for two values of Lorentz factors along $\hat{z}$ at the jet axis ($\Gamma_{c,j}$).}
\label{tab:initial_param}
\begin{tabular}{lcccccc}
\hline
\hline
Run ID & $z$ & $B_{zc,j}$ & $\Omega_c$ & $\overline{P_{0,j}}$ & $N_e$ & $t_{\rm final}$ \\
    &     &  $\times 10^{-5}$   &  $\times 10^{-2}$   &   $\times 10^{-05}$  &   $\times 10^{-9}$  &  $\times 10^{3}$   \\
(1) & (2) & (3) & (4) & (5) & (6) & (7) \\
\hline
\multicolumn{7}{l}{$\Gamma_{c,j}=5$} \\
Rg5z0p1 & 0.1 & 4.003 & 2.25 & 3.86 & 1.747 & 3.5 \\
Rg5z2   & 2.0 & 4.003 & 2.25 & 3.86 & 1.747 & 3.5 \\
Rg5z5   & 5.0 & 4.003 & 2.25 & 3.86 & 1.747 & 3.5 \\
Rg5z6   & 6.0 & 4.003 & 2.25 & 3.86 & 1.747 & 3.5 \\
\hline
\multicolumn{7}{l}{$\Gamma_{c,j}=2$} \\
Rg2z0p1 & 0.1 & 1.575 & 5.56 & 0.4 & 0.134 & 8.0 \\
Rg2z2   & 2.0 & 1.575 & 5.56 & 0.4 & 0.134 & 8.0 \\
Rg2z5   & 5.0 & 1.575 & 5.56 & 0.4 & 0.134 & 8.0 \\
\hline
\end{tabular}
\end{table}

\section{Hybrid Eulerian-Lagrangian Particle Framework}\label{sec:3}
We employ the hybrid Eulerian-Lagrangian particle module of the \textsc{PLUTO} code \citep{Vaidya2018, Mukherjee2021} to model the non-thermal emission from relativistic jets. In this approach, Lagrangian macro-particles are injected at the jet nozzle at fixed time intervals and advected with the flow, sharing the bulk velocity of the surrounding plasma. Each particle represents an ensemble of relativistic micro-particles (electrons) characterized by a time-dependent energy distribution $N(\gamma, \tau)$, where $\gamma$ is the micro-particle Lorentz factor and $\tau$ is the proper time in the jet comoving frame.

Particles are initialized within a circular cross-section of radius $r_j$ located just above the injection nozzle. The injection region is divided into 20 radial bins and 36 angular bins spanning $0 < \phi < 2\pi$, resulting in 72 particle injection locations per injection cycle. Over the course of the simulations, this procedure yields approximately $1.5 \times 10^{6}$ particles in the Rg5 runs and about $2 \times 10^{6}$ particles in the Rg2 runs, providing sufficient sampling of the evolving jet volume. At injection, the micro-particle population follows a power-law distribution in Lorentz factor,
\begin{equation}
\label{eq:power-law}
    N(\gamma, \tau=0) = N_0 \, \gamma^{-\alpha},
\end{equation}
with a steep slope $\alpha = 6$ and cutoffs at $\gamma_{\rm min} = 10^2$ and $\gamma_{\rm max} = 10^8$. The normalization $N_0$ is fixed by requiring
\begin{equation}
\label{eq:nmicro}
    \int_{\gamma_{\rm min}}^{\gamma_{\rm max}} N_0 \, \gamma^{-\alpha} \, d\gamma = N_{\rm e}\,,
\end{equation}
where $N_{\rm e}$ is the electron number density (Table~\ref{tab:initial_param}).
We estimate $N_{\rm e}$ under the assumption that the energy density of relativistic electrons, $U_{\rm e}$, constitutes a fraction of the magnetic field energy density in the injection region,
\begin{equation}
\label{eq:equip}
    U_{\rm e} = m_e c^2 \int_{\gamma_{\rm min}}^{\gamma_{\rm max}} \gamma \, N(\gamma) \, d\gamma 
    =  \epsilon \frac{B_{\rm dyn}^2}{8 \pi}\,,
\end{equation}
where $B_{\rm dyn}$ is the average magnetic field strength at injection and $\epsilon$ specifies the fractional ratio, fixed to $2 \times 10^{-4}$ for all simulations. Combining Equations~\ref{eq:power-law}-\ref{eq:equip} yields
\begin{equation}
    N_{\rm e} =  \frac{\epsilon}{m_e c^2} \frac{B_{\rm dyn}^2}{2} \left(\frac{2-\alpha}{1-\alpha}\right)
    \left( \frac{\gamma_{\rm max}^{1-\alpha} - \gamma_{\rm min}^{1-\alpha}}{\gamma_{\rm max}^{2-\alpha} - \gamma_{\rm min}^{2-\alpha}} \right).
\end{equation}
After injection, the spectral evolution of each particle is governed by the Fokker-Planck equation, \citep{Vaidya2018}
\begin{equation}
    \frac{\partial \chi}{\partial \tau} +
    \frac{\partial}{\partial \gamma}
    \left[ \left( -\frac{\gamma}{3} \nabla_\mu u^{\mu} + \dot{\gamma}_{\mathrm{rad}} \right) \chi \right] = 0,
    \label{eq:fokker-planck}
\end{equation}
where $\chi = N/n$ is the spectrum normalized to the local fluid number density $n$, $\tau$ is the proper time in the jet comoving frame and $u^\mu$ is the bulk four-velocity. The first term inside the brackets represents adiabatic losses; using the continuity equation $\nabla_{\mu}(n u^{\mu}) = 0$ gives $\nabla_\mu u^{\mu} = (1/\rho)(d\rho/d\tau)$. The second term, $\dot{\gamma}_{\mathrm{rad}}$, accounts for radiative losses; in this work, we consider only synchrotron and IC processes. During their evolution, these particles may encounter shocks and are consequently accelerated as a result of Diffusive Shock Acceleration \citep{Vaidya2018, Mukherjee2021, Dubey2023}.

\subsection{Radiative Losses} \label{sec:radiative_losses}
Relativistic electrons in astrophysical jets cool primarily via synchrotron radiation and IC scattering. The total radiative cooling rate for an electron with Lorentz factor $\gamma$ in the comoving frame can be expressed as
\begin{equation}
\left( \frac{d\gamma}{d\tau} \right)_{\mathrm{rad}} =
\left( \frac{d\gamma}{d\tau} \right)_{\mathrm{sync}} +
\left( \frac{d\gamma}{d\tau} \right)_{\mathrm{IC}}.
\end{equation}
The synchrotron cooling rate for an ultra-relativistic electron spiraling in a magnetic field depends on the comoving magnetic energy density $ U_B = B^2 / (8\pi) $ and scales with the square of the Lorentz factor,
\begin{equation}
\left( \frac{d\gamma}{d\tau} \right)_{\mathrm{sync}} =
- \frac{4}{3} \, \frac{\sigma_T c}{m_e c^2} \, \gamma^2 \beta^2 U_B,
\end{equation}
with $\sigma_T$ the Thomson cross-section, $m_e$ the electron mass, $c$ the speed of light, and $\beta \equiv v/c \simeq 1$ for the highly relativistic electrons considered here.

IC losses arise from the scattering of ambient photons to high-energy particles. In this work, we consider only the CMB as the target photon field, modelled as an isotropic blackbody of temperature $T_{\mathrm{CMB}}$. In the comoving frame, the CMB energy density is
\begin{equation}
U_{\mathrm{CMB}} = a_{\mathrm{rad}} T_{\mathrm{CMB}}^4 (1+z)^4 \Gamma^2 \left( 1 + \frac{\beta^2}{3} \right)
\end{equation}
\citep{Dermer2009},
where $a_{\mathrm{rad}}$ is the radiation constant, $z$ is the source redshift, $\Gamma$ is the bulk Lorentz factor of the jet, and $\beta$ is the dimensionless bulk velocity. The corresponding IC cooling rate is
\begin{equation}
\left( \frac{d\gamma}{d\tau} \right)_{\mathrm{IC}} =
- \frac{4}{3} \, \frac{\sigma_T c}{m_e c^2} \, \gamma^2 U_{\mathrm{CMB}} \,
\frac{\gamma_K^2}{\gamma^2 + \gamma_K^2}
\end{equation}
\citep{Schlickeiser2010}, where the Klein–Nishina (KN) transition occurs at the critical Lorentz factor
\begin{equation}
\gamma_K = \frac{3\sqrt5 }{8\pi}\frac{ m_e c^2}{k_B T_{\mathrm{CMB}} (1+z)}.
\end{equation}
For Lorentz factors $\gamma > \gamma_{K}$, the full KN cross-section has to be used resulting in a significant reduction of the IC loss rate. 
Assuming a present-day CMB temperature of $T_{\mathrm{CMB}} = 2.728~\mathrm{K}$ at $z = 0$, we obtain $\gamma_K \simeq 5.9 \times 10^8 (1+z)^{-1}$, 
indicating that KN effects becomes significant only for extremely high-energy electrons.

Combining the synchrotron and IC contributions, the total radiative cooling rate is
\begin{equation}
\left( \frac{d\gamma}{d\tau} \right)_{\mathrm{rad}} =
- \frac{4}{3} \, \frac{\sigma_T c}{m_e c^2} \, \gamma^2
\left[ U_B + U_{\mathrm{CMB}} \,
\frac{\gamma_K^2}{\gamma^2 + \gamma_K^2} \right].
\label{Cooling Equation}
\end{equation}
\subsection{Emission Signatures}
This section describes the modified emissivity module implemented within the particle framework of the PLUTO code. We employ two approaches to compute the observed flux density for synchrotron and IC/CMB emission. For synchrotron radiation, we first evaluate the emissivity in the comoving frame and then transform it to the observer’s frame. For the IC emission from CMB seed photons, we adopt the formalism of \citet{Georganopoulos2001}.

Hereafter, quantities in the observer’s frame are denoted without primes, those in the comoving frame of a macro-particle with primes, and those in the AGN (cosmological rest) frame at redshift $z$ with a hat.
\subsubsection{Synchrotron Emission} \label{sec:synchrotron}
In our calculations, we assume that the electrons within each macro-particle are isotropically distributed, such that all pitch angles are equally probable. To account for this in our simulation, we adopt the formalism of \citet{Crusius1986} to compute the synchrotron emissivity,
%
\begin{equation}\label{synchrotron_emissivity_eq}
\epsilon'_s J'_{\mathrm{sync}}(\epsilon'_s) = \frac{\sqrt{3} e^3 \epsilon'_s}{4\pi m_e c^2} B' \int d\gamma' \, N(\gamma') \, R(x),
\end{equation}
where $B'$ is the comoving magnetic-field strength, $N(\gamma')$ is the electron number density, $\gamma'$ is the Lorentz factor of the micro-particle, and $\epsilon'_s = h \nu_{\rm obs}(1+z) / (m_e c^2 \delta_D)$ is the dimensionless photon energy and the doppler factor is defined as usually,
\begin{equation}
\delta_D = \frac{1}{\Gamma_p \left( 1 - \beta \cos(\theta_{\rm obs}) \right)}\,,
\end{equation}
where $\Gamma_p$ and $\beta_p = v_p/c$ are the bulk Lorentz factor and the velocity normalized to light speed of the macro-particle,
and $\theta_{\rm obs}$ is the angle between the particle’s velocity vector and the observer’s line of sight. 
The dimensionless parameter $x$ is defined as
\begin{equation}
x = \frac{4\pi \epsilon'_s m_e^2 c^3}{3 e B' h \gamma'^2}.
\end{equation}
The function $ R(x) $ represents the pitch-angle-averaged synchrotron kernel, defined as
\begin{equation}
R(x) = \frac{x}{2} \int_0^{\pi} d\theta \, \sin\theta \int_{x/\sin\theta}^{\infty} dt \, K_{5/3}(t),
\end{equation}
where $ K_{5/3} $ is the modified Bessel function of order $5/3$.

For the numerical implementation, we adopt the analytic approximation for $ R(x) $ as provided by \citet{Aharonian2010},
\begin{equation}
R(x) \approx \frac{1.808 \, x^{1/3}}{\sqrt{1 + 3.4 \, x^{2/3}}}  \frac{1 + 2.21 \, x^{2/3} + 0.347 \, x^{4/3}}{1 + 1.353 \, x^{2/3} + 0.217 \, x^{4/3}} \, e^{-x}.
\end{equation}
The observed flux density is defined as
\begin{equation} \label{vFv}
\nu F_\nu = \delta_D^4 V'_b \frac{\epsilon'_s J'(\epsilon'_s)}{4\pi d_L^2}
\end{equation}
\citep{Dermer2009, Finke2008},
where $V'_b$ is the comoving volume of the emitting region and $d_L$ is the luminosity distance to the source.

\subsubsection{Inverse Compton Scattering of CMB Photons (IC/CMB)}\label{sec:IC_CMB}
In this work, we extend the particle module by incorporating a full calculation of IC/CMB emissivity, including the effects of relativistic transformations and the Klein–Nishina suppression at high electron energies. This addition enables a consistent computation of the high-energy emission from relativistic jets within the same framework used for synchrotron radiation.

We approximate the CMB photon field as isotropic and monoenergetic, with the photon energy taken to be the mean energy of a blackbody spectrum. 
In the AGN (cosmological rest) frame at redshift $z$, the dimensionless CMB photon energy is given by
\begin{equation}\label{CMB_Eng}
\hat{\epsilon}_* = \frac{2.70\, k_B T_{\mathrm{CMB}} (1 + z)}{m_e c^2}.
\end{equation} 
The energy density of the isotropic CMB radiation field in the AGN frame is
\begin{equation}
\hat{U}_{iso} = a \left[ T_{\mathrm{CMB}}(1 + z) \right]^4,
\end{equation}
where $a$ is the Stefan-Boltzmann constant.
To compute the IC observed flux density from CMB seed photons, we follow the formalism of \citet{Georganopoulos2001}. 
The observed $\nu F_{\nu}$ flux is
\begin{multline}\label{JnuIC_big}
    \nu F_\nu = \frac{3}{4} \frac{\sigma_T c \hat{\epsilon}^2_s}{4\pi d_L^2}\delta_D^3 V'_b \int_0^\infty d\hat{\epsilon}_* \, \frac{\hat{u}_*(\hat{\epsilon}_*)}{\hat{\epsilon}_*^2} \\ \int_{\hat{\gamma}_{\min}}^{\hat{\gamma}_{\max}} d\hat{\gamma} \, \frac{N_e(\hat{\gamma})}{\hat{\gamma}^2} \, \hat{F_C}(\hat{q}, \hat{\Gamma}_e),
\end{multline}
where $ \hat{\epsilon}_s = h\nu_{obs}(1+z)/ m_e c^2$ is the dimensionless scattered photon energy, 
$\hat{u}_*(\hat{\epsilon}_*) $ is the external isotropic radiation field, 
${V'_b}$ is the comoving volume of the emitting region, and 
$ N_e(\hat{\gamma}) $ is isotropic electron number density per unit Lorentz factor $\hat{\gamma} = \delta_D \gamma '$. 
The Compton kernel $ \hat{F_C}(\hat{q}, \hat{\Gamma}_e) $ for isotropic photon and electron distribution is
\begin{multline}
\hat{F_C}(\hat{q}, \hat{\Gamma}_e) =
\left[ 2\hat{q} \ln \hat{q} + (1 + 2\hat{q})(1 - \hat{q}) \right. \\
\left. + \frac{1}{2} \frac{(\hat{\Gamma}_e \hat{q})^2}{1 + \hat{\Gamma}_e \hat{q}} (1 - \hat{q}) \right]
H\left(\hat{q}; \frac{1}{\hat{\gamma}^2}, 1\right)
\end{multline}
\citep{Jones1968,BLUEMENTHAL1970},
where $ H(\hat{q}; a, b) $ is a Heaviside step function,
\begin{equation}
H(\hat{q}; a, b) =
\begin{cases}
1, & \text{if } a \leq \hat{q} \leq b, \\
0, & \text{otherwise}.
\end{cases}
\end{equation}
while
\begin{align}
\hat{q} &= \frac{\hat{\epsilon}_s / \hat{\gamma}}{\hat{\Gamma}_e (1 - \hat{\epsilon}_s / \hat{\gamma})}, \\
\hat{\Gamma}_e &= 4 \hat{\epsilon}_* \hat{\gamma}.
\end{align}
The limits on $\hat{q}$ are
\begin{equation}\label{conditon on q}
    \frac{1}{4\hat{\gamma}^2}\leq \hat{q} \leq 1,
\end{equation}
which in turn implies limits for the integration over $\hat{\gamma}$,
\begin{equation}
\hat{\gamma}_{\min}
= \frac{1}{2}\,\hat{\epsilon}_s
\left(
1 + \sqrt{1 + \frac{1}{\hat{\epsilon}_*\,\hat{\epsilon}_s}}
\right),
\end{equation}

\begin{equation}
\hat{\gamma}_{\max}
= \frac{\hat{\epsilon}_*\,\hat{\epsilon}_s}{\hat{\epsilon}_* - \hat{\epsilon}_s}\,
H\!\left(\hat{\epsilon}_* - \hat{\epsilon}_s\right)
+ \hat{\gamma}_2\,
H\!\left(\hat{\epsilon}_s - \hat{\epsilon}_*\right).
\end{equation}
Considering equations~\ref{CMB_Eng} to~\ref{conditon on q}, equation~\ref{JnuIC_big} reduces to

\begin{align}\label{IC_emissivity_eq}
\nu F_\nu = 
&\frac{3}{4} \frac{\sigma_T c}{4\pi d_L^2}\delta_D^3 V'_b \hat{U}_{iso} 
\left( \frac{\hat{\epsilon}_s}{\hat{\epsilon}_*} \right)^2 \notag \\
&\times \int_{\hat{\gamma}_{\min}}^{\hat{\gamma}_{\max}} d\hat{\gamma} \, \frac{N_e(\hat{\gamma})}{\hat{\gamma}^2} \, \hat{F_C}(\hat{q}, \hat{\Gamma}_e)
\end{align}
A detailed comparison between the spectral energy distributions (SEDs) generated by the updated particle module of the PLUTO code and those obtained with the open-source Python package AGNpy \citep{Nigro2022} is 
presented in Appendix~\ref{Appendix 2}, providing a successful test of our implementation.

\section{Jet Dynamics and Synthetic Emission}\label{sec:4}
This section describes the main results of the study, focusing on dynamical signatures from 3D RMHD simulations along with associated synthetic emission signatures with varying redshift.

\begin{figure*}
    \centering
    \includegraphics[width=0.8\linewidth]{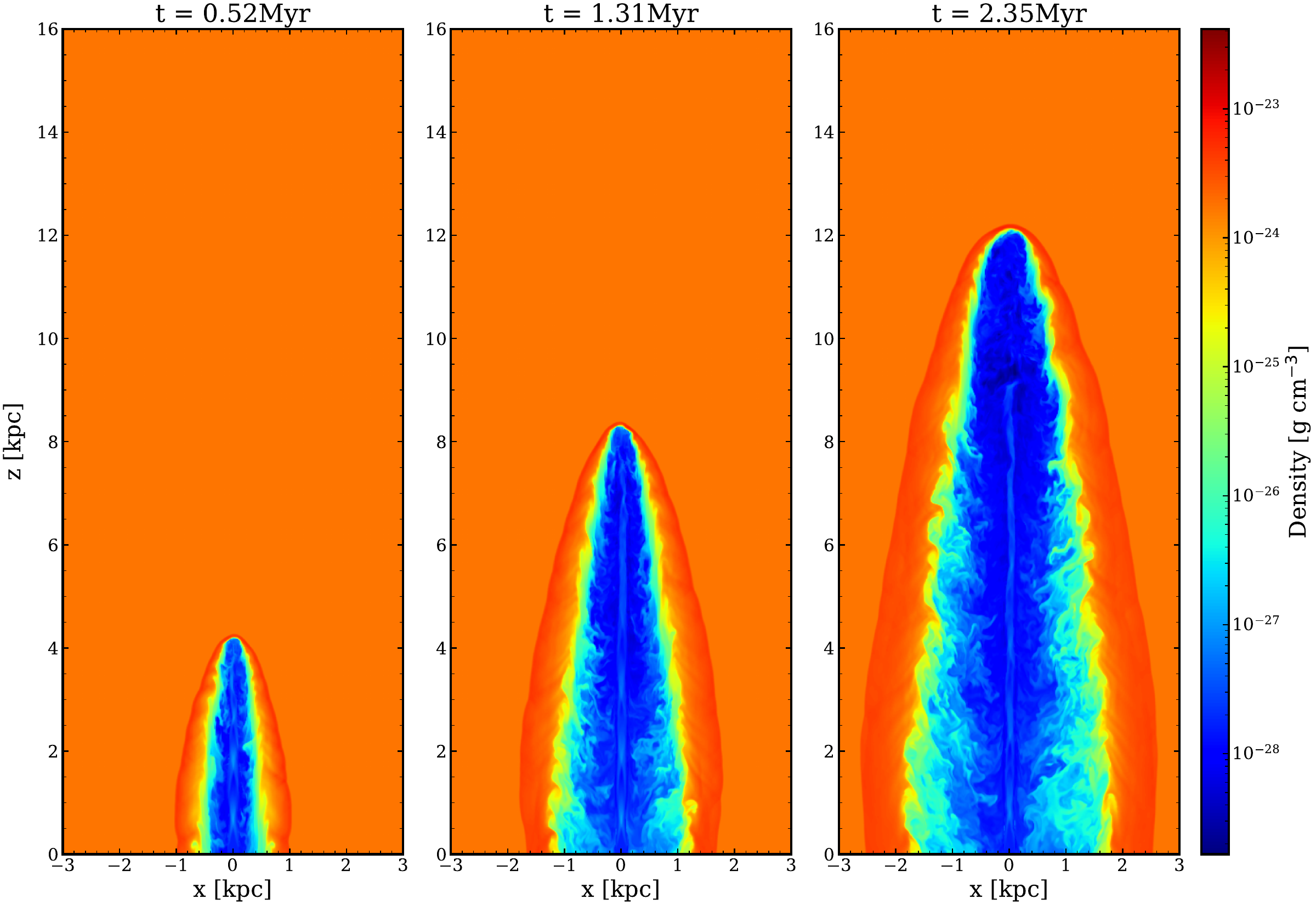}
    \includegraphics[width=0.8\linewidth]{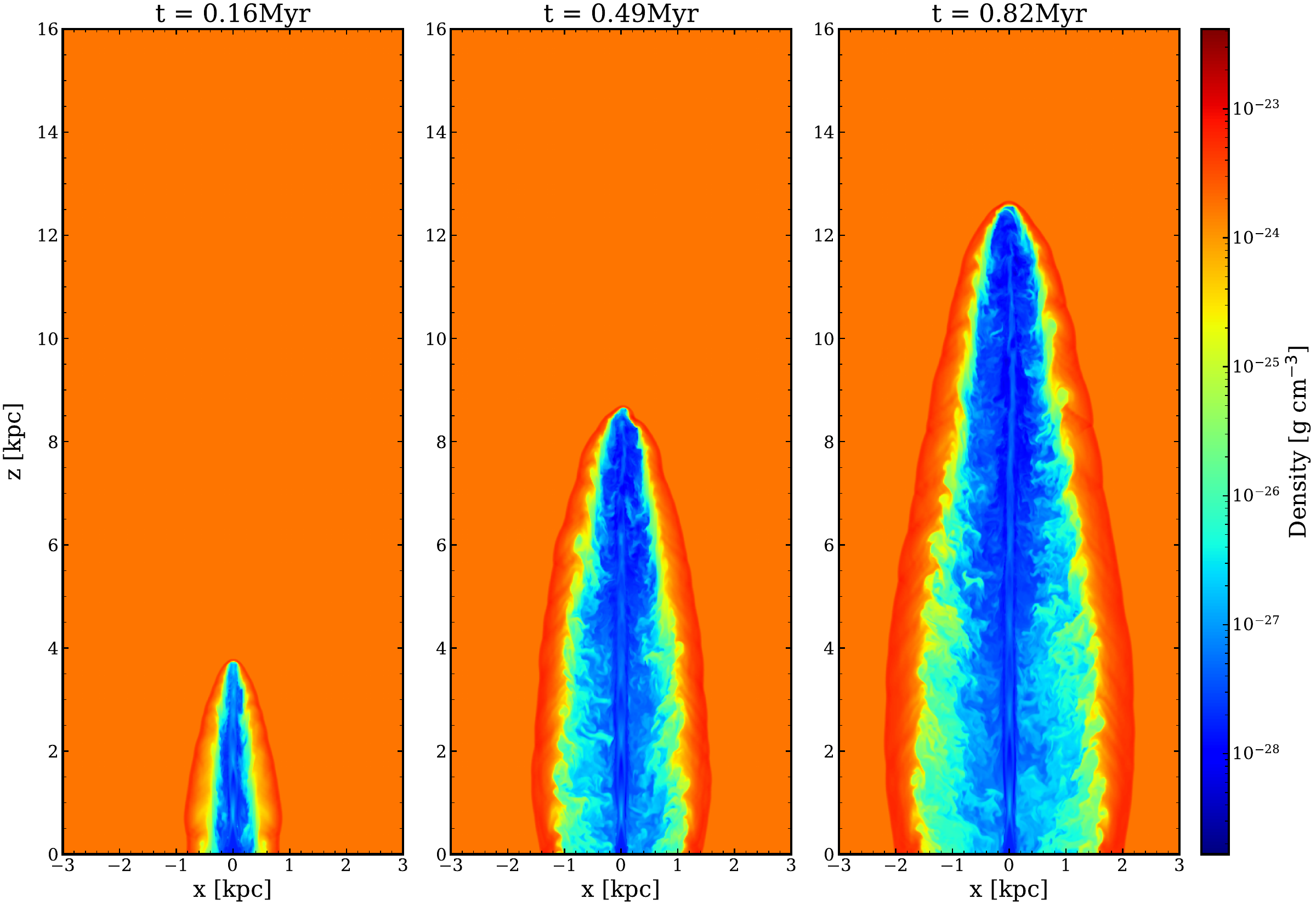}
    \caption{Density slices in the x–z plane at y = 0 showing the time evolution of the jet in the Rg2 (above) and Rg5 (below) simulation at selected time steps. The logarithmic color scale represents density in cgs units. The jet propagates along the z-axis, exhibiting morphological changes, internal structure formation, and interaction with the ambient medium over time.    }
    \label{Dynamical Evolution of jet}
\end{figure*}

\subsection{Jet Dynamical Evolution}
Figure~\ref{Dynamical Evolution of jet} shows the dynamical evolution of the jets for models Rg2 and Rg5. 
The adopted Lorentz factors, $\Gamma=2$ and 5, are motivated by several observational and simulation studies of powerful FR~II/blazar jets, 
which show that the flow can remain mildly to moderately relativistic even on kpc scales \citep[e.g.,][]{2011ApJ...730...92H, Mukherjee2020, Upreti2024}.
In both simulations, the jet propagates in the $z$-direction at relativistic speed, driving a strong bow shock into the ambient medium and thereby creating a cocoon of shocked, low-density gas surrounding the jet beam. 

The inner spine of the jet close to its nozzle show formation of re-collimation shocks as evident from the strong periodic density enhancement \citep{2008A&A...486..663K, 2018A&A...609A.122B, Dubey2023, 2023JPlPh..89e9101P}. The central spine at larger length scales (along the z axis) loses its symmetry due growth of pressure driven instability resulting in wiggle and ultimately breaking into turbulent flow beyond $10\,\mathrm{kpc}$ in the top right panel for Rg2 simulation.

As the jet evolves, the turbulent cocoon expands laterally, additionally the backflow material from the jet head interacts with the central jet spine and outer shocked ambient medium. The shear that exists between thes super-Alfvenic flows result in formation of Kelvin Helmholtz \citep{2021A&A...649A.150B} driven vortices as evident in the figure (green and yellow colors).

The time steps shown for each model correspond to stages at which these jets reach similar physical extent, allowing a direct comparison of their morphology.
For the Rg2 jet, the snapshots at $t = 0.52$, 1.31, and 2.35 Myr capture deprojected jet lengths of approximately 4, 8, and 12 kpc, respectively. 
The Rg5 jet reaches the same distance much earlier, at $t = 0.16$, 0.49, and 0.82 Myr, highlighting its more rapid advance through the ambient medium. 
At these matched spatial scales, the lower-Lorentz-factor jet exhibits a wider cocoon, stronger lateral expansion, and greater deformation along the beam–cocoon interface. In contrast, the higher-Lorentz-factor jet retains tighter collimation and produces a thinner cocoon. These differences underscore the influence of the jet’s initial Lorentz factor on its overall morphology and rate of propagation.

The structure -- density and pressure distribution -- of the ambient medium is expected to evolve with redshift.
For example, galaxies at higher redshift are likely to host a denser and more turbulent gaseous environments \citep{2025MNRAS.541.1707T, 2025MNRAS.540.3350A}. 
Such variations may significantly influence the morphology and evolution of relativistic jets. 
In particular, a denser and more inhomogeneous ambient medium can enhance the jet confinement and also amplify the jet–ambient interaction, thereby promoting turbulence, entrainment, and jet deceleration, while also affecting the long-term stability and propagation of the jet \citep{2025Galax..13..102M}.

For the present work, however, we decided to keep all ambient medium parameters fixed within a given class of simulations. 
This allows us too ensure that any differences in the resulting emission features are indeed exclusively arising from the redshift-dependent radiative
processes discussed below, rather than from variations in jet dynamics or external environmental conditions.

Our assumption is thus not a limitation, but is necessary for being able to disentangle the physical processes at work.
A future approach may then investigate and compare the additional impact by environmental effects. 
We note that these effects are not only an issue connected to the global redshift evolution, 
but also a local effect, meaning that different jet sources naturally reside in a different local environment.

\subsection{Synthetic Emission}\label{Synthetic emission}
\begin{figure*}[t]
    \centering
    \includegraphics[width=1.0\linewidth]{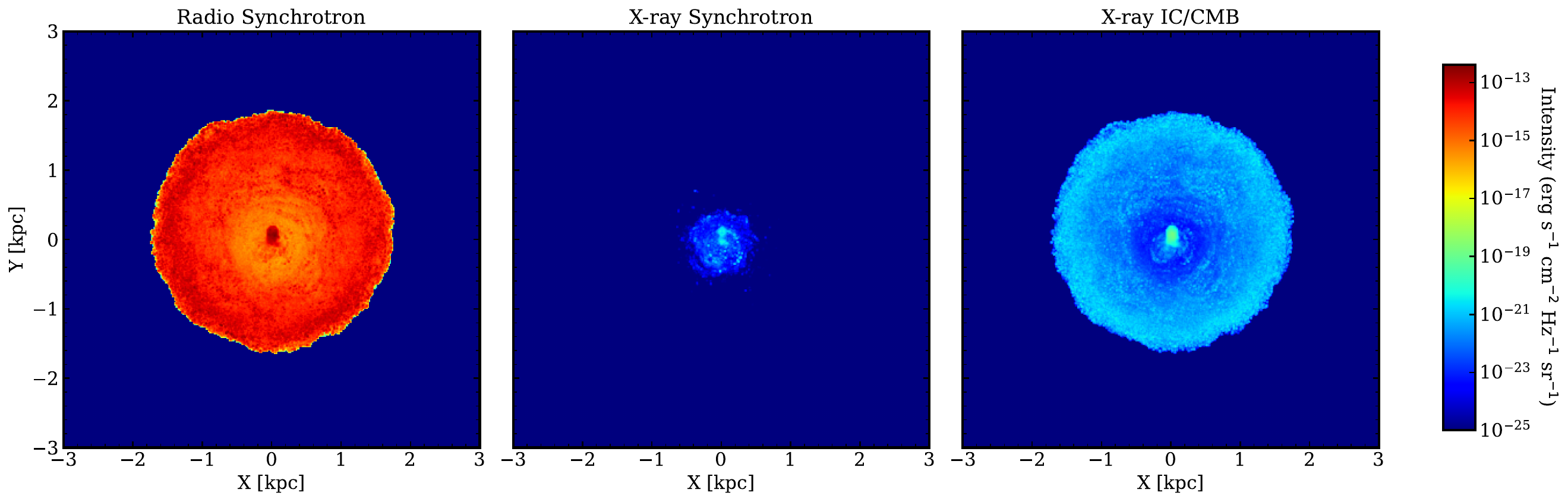}
    \caption{Intensity maps for Rg5z5 simulation at time 0.82 Myr for a viewing angle of $1^\circ$.
    \textbf{Left:} Radio synchrotron emission at observed frequency 500 MHz showing contributions from both the jet axis and cocoon. 
    \textbf{Center:} X-ray synchrotron emission at 1~keV, confined to regions with very high-energy electrons along the jet axis. 
    \textbf{Right:} X-ray IC/CMB emission at 1~keV, tracing the full jet due to interactions with the cosmic microwave background.}
    \label{Radiative Intensity}
\end{figure*}

To connect the dynamical evolution of relativistic jets with their observable signatures across redshift, we compute the synthetic emission produced by the non-thermal electron population for the Rg5z5 simulation at time 0.82 Myr.
We integrate the observed emissivity along our line of sight (LOS) direction, assuming that there is no absorption of light as it traverses the medium (hence assuming optical depth $\tau$ = 0), to produce synthetic two-dimensional intensity map into the plane of the sky orthogonal to the LOS (Figure \ref{Radiative Intensity}). 
All emissions are calculated for a viewing angle of $1^\circ$.

The emitted frequency of radiation depends on both the electron energy distribution and the energy density of background fields, such as the magnetic field and the CMB photon field. For synchrotron radiation, the scattered frequency is determined by the electron Lorentz factor $\gamma$ and the local magnetic field strength $B$,
\begin{equation}
    \nu_{\rm sync} \propto \gamma^2 B
\end{equation}
\citep{BLUEMENTHAL1970}.
As a result electrons with moderate Lorentz factors predominantly emit in the radio band throughout the jet (Figure~\ref{Radiative Intensity}, left panel). 
On the other hand synchrotron X-rays require highly energetic electrons with $\gamma \sim 10^8$ or higher \citep{Rahman2023, Ighina2022}. 
In our simulations, these ultra-relativistic electrons are confined along the jet axis, where efficient acceleration happens, leading to synchrotron X-ray
emission that is strongly localized along the jet axis (Figure~\ref{Radiative Intensity}, middle panel).

In contrast, for the IC/CMB process, the scattered frequency depends on the electron Lorentz factor and the average CMB photon energy $\epsilon_{\rm CMB}$,
\begin{equation}
    \nu_{\rm IC} \propto \gamma^2 \epsilon_{\rm CMB}
\end{equation}
\citep{BLUEMENTHAL1970}.
At redshift $z$, the CMB energy density increases as $(1+z)^4$, and the typical photon energy scales as $(1+z)$. Consequently, electrons with moderate Lorentz factors can upscatter CMB photons into the X-ray band at high redshift, making IC/CMB emission an efficient mechanism throughout large-scale jets \citep{Ghisellini2014}, as shown in Figure~\ref{Radiative Intensity} (right panel).
A significant portion of the IC/CMB emission comes from the cocoon and backflow regions \citep{Nath2010}, where relativistic electrons accumulate over time and experience comparatively weak re-acceleration.

\subsection{Synthetic Spectral Energy Distribution}
Figure~\ref{Radiative Intensity} highlights the spatial regions contributing to the radiative emission. To examine how the observed flux density varies across different wavebands, we compute synthetic SEDs for both synchrotron and IC/CMB processes (Equations~\ref{vFv}, ~\ref{synchrotron_emissivity_eq} and \ref{IC_emissivity_eq}). 

The synchrotron and IC/CMB SEDs are evaluated at selected frequencies\footnote{Synchrotron: $10^{5}$, $10^{6}$, $10^{7}$, $5\times10^{7}$, $10^{8}$, $5\times10^{8}$, $5\times10^{9}$, $5\times10^{10}$, $5\times10^{12}$, $2.418\times10^{16}$, $2.418\times10^{17}$, $2.418\times10^{18}$~Hz.
IC/CMB: $5\times10^{14}$, $2.418\times10^{16}$, $2.418\times10^{17}$, $2.418\times10^{18}$, $2.418\times10^{19}$, $2.418\times10^{20}$, $2.418\times10^{22}$, $2.418\times10^{24}$~Hz.} for all simulation runs.

In this section, we investigate the physical processes responsible for the X-ray emission at different stages of jet evolution and across cosmic time. 
To distinguish between synchrotron and IC/CMB contributions at a given evolutionary stage, we construct synthetic SEDs covering the relevant observational frequency range and examine the relative importance of each emission component.

\begin{figure}
    \centering
    \includegraphics[width=\linewidth]{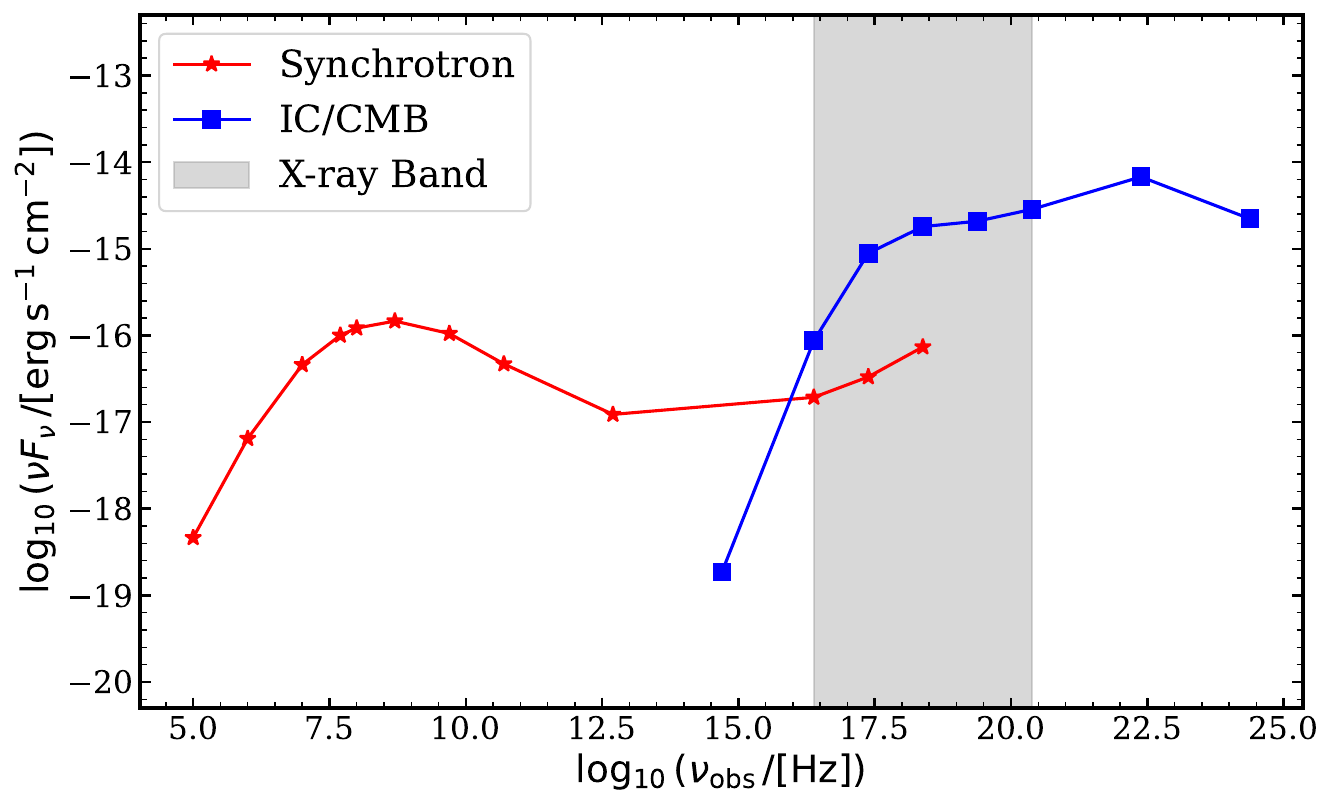}
    \caption{SED for simulation run \texttt{Rg5z5} at time step $t = 0.82$~Myr. The synchrotron component is shown by red star symbols, whereas the IC/CMB component is shown by blue square symbols. The shaded gray region indicates the X-ray band ($0.1$-$1000$~keV).}
    \label{Net SED}
\end{figure}

As a representative case, we adopt the Rg5z5 simulation as a reference model and examine its SED at $t = 0.82$~Myr.
Figure~\ref{Net SED} shows the characteristic double-humped SED structure, with the low-energy hump arising from synchrotron emission (red stars) and the high-energy hump produced by the IC/CMB component (blue squares).
The shaded gray region, spanning frequencies from $ 2.418\times10^{16}$ to $2.418\times10^{20}$~Hz, denotes the X-ray band ($0.1$-$1000$~keV). 
At this time step and redshift, the X-ray emission is dominated by the IC/CMB component, with a comparatively weaker contribution from synchrotron radiation.

\begin{figure*}
    \centering
    \includegraphics[width=1\linewidth]{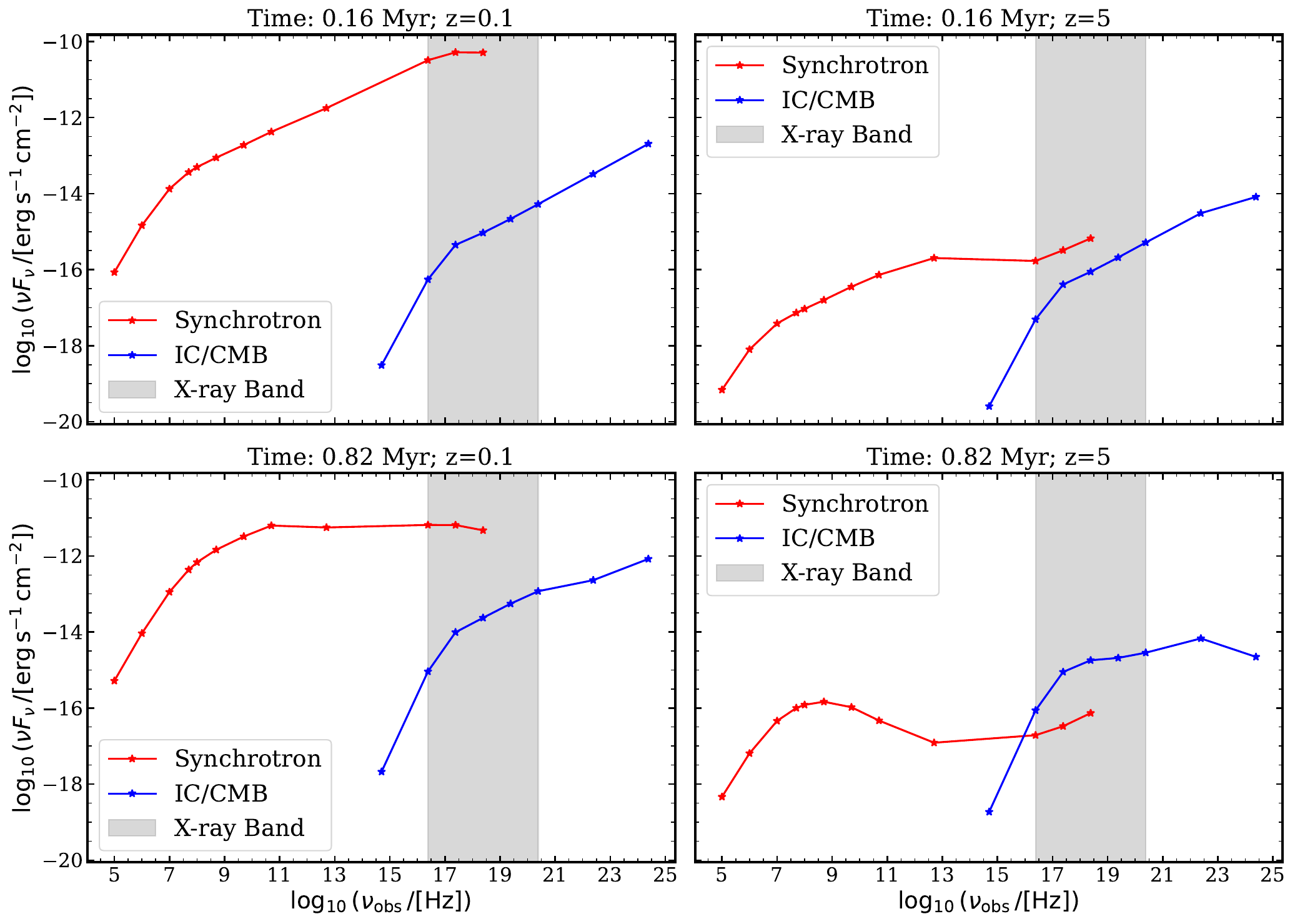}
    \caption{SEDs of the Rg5 simulation at two time steps (rows) and two redshifts (columns). 
    Each panel displays the synchrotron (red) and IC/CMB (blue) components as functions of frequency. The top and bottom rows correspond to simulation times of $t = 0.16$~Myr and $t = 0.82$~Myr, respectively, while the left and right columns represent redshifts $z = 0.1$ and $z = 5$. The gray shaded region denotes the X-ray band ($0.1-1000$~keV).}
    \label{fig:SED_time_z_grid}
\end{figure*}

To assess the dependence of the X-ray emission on both redshift and jet evolution, we compare the SEDs at two redshifts ($z = 0.1$ and $z = 5$) and two evolutionary stages ($t = 0.16$~Myr and $t = 0.82$~Myr) for Rg5 simulations, as shown in Figure~\ref{fig:SED_time_z_grid}. 

At the early evolutionary stage ($t = 0.16$~Myr), the X-ray band is primarily dominated by synchrotron emission at both redshifts (Figure~\ref{fig:SED_time_z_grid}, top panels). At this stage the jet still contains a substantial population of freshly accelerated high-energy electrons capable of reaching Lorentz factors high enough to produce X-ray emission through synchrotron radiation. Increasing the redshift from $z=0.1$ to $z=5$ leads to a noticeable reduction in the synchrotron flux. This behavior arises from the rapid increase of the CMB energy density with redshift, $U_{\rm CMB} \propto (1+z)^4$, which enhances IC cooling of the highest-energy electrons. As a result, electrons lose energy more efficiently through IC/CMB scattering, suppressing the high-energy tail of the electron distribution and reducing the synchrotron X-ray emission. Nevertheless, when the jet is young and compact the population of energetic electrons remains sufficiently large that synchrotron radiation can still contribute significantly to the X-ray band even at high redshift.

At later evolutionary stages ($t = 0.82$~Myr), the electron population undergoes significant radiative cooling, which strongly affects the high-energy end of the distribution. The enhanced IC/CMB losses at high redshift efficiently cool the most energetic electrons, preventing a substantial population from reaching Lorentz factors required to produce synchrotron X-ray emission. Consequently, at low redshift ($z = 0.1$) synchrotron radiation can still dominate the X-ray band due to the presence of a larger population of high-energy electrons. In contrast, at high redshift ($z = 5$) the increased CMB energy density both suppresses the synchrotron X-ray emission and boosts the IC/CMB component, leading to IC/CMB-dominated X-ray emission (Figure~\ref{fig:SED_time_z_grid}, bottom panels).

These results indicate that both synchrotron and IC/CMB processes can produce X-ray emission even at high redshift, but the dominant mechanism depends strongly on the evolutionary stage and physical size of the jet. Compact and young jets are more likely to produce X-rays through synchrotron radiation because they still contain a substantial population of high-energy electrons. In contrast, larger and more evolved kpc-scale jets tend to produce X-rays predominantly through IC/CMB scattering, as radiative cooling reduces the high-energy electron population while the CMB energy density becomes increasingly important at high redshift.

\subsection{Dependence of radio spectral index on redshift: the \texorpdfstring{$\alpha$-$z$}{alpha-z} relation}
Along with the enhancement of X-ray flux with redshift we also observe variation in the radio flux and spectral index as a function of redshift.
The radio spectral index $\alpha$ and the flux density are related as
\begin{equation}
\log(F_\nu) = \alpha \, \log(\nu) + \mathrm{constant},
\label{eq:alphaz}
\end{equation}
To determine the spectral index in our simulation, we compute $\log(F_\nu)$ versus $\log(\nu)$ at four radio frequencies: 0.1, 0.5, 5.0, and 50~GHz. 
We then perform a linear least-squares fit using the Eq.~\ref{eq:alphaz}, such that the fitted slope directly yields $\alpha$ for each combination of evolutionary time and redshift.

\begin{figure*}
\centering
\includegraphics[width=1\linewidth]{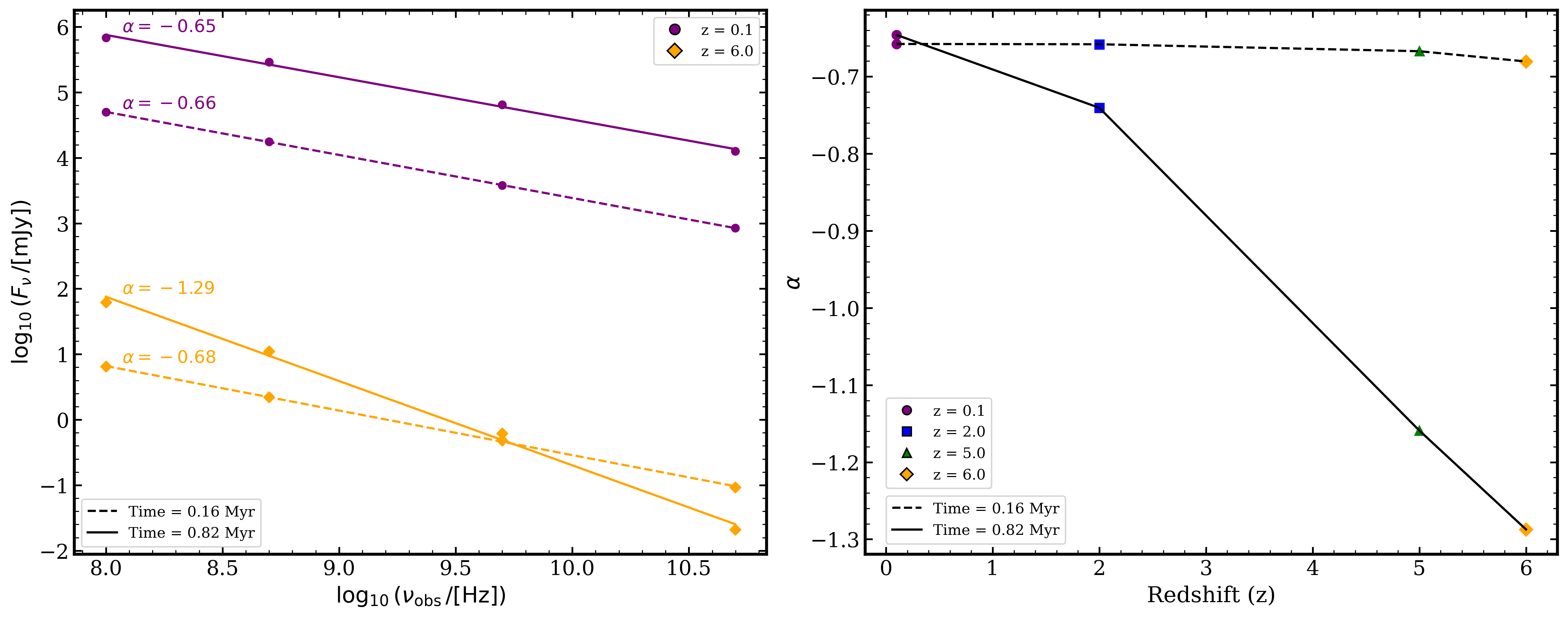}
\caption{Left: Log-log radio spectra used to compute the spectral index $\alpha$ for the Rg5 simulations at two evolutionary times, $t = 0.16$~Myr (dashed lines) and $t = 0.82$~Myr (solid lines). Filled circles and diamonds show the simulated flux densities at observing frequencies of 0.1, 0.5, 5.0, and 50~GHz for $z = 0.1$ and $z = 6.0$ respectively. Lines indicate the best-fitting linear relations of the form $\log(F_\nu) = \alpha \log(\nu) + \mathrm{constant}$, with the fitted slopes ($\alpha$) annotated. The spectra steepen with time for the $z > 0.1$. 
Right: Spectral index $\alpha$ as a function of redshift for the same Rg5 simulations at $t = 0.16$~Myr (dashed line) and $t = 0.82$~Myr (solid line), derived from the fits shown in the left panel. 
At early times, $\alpha$ shows only a weak dependence on redshift, whereas at later times the spectra steepen rapidly with increasing $z$, illustrating the emergence of a pronounced $\alpha$-$z$ relation in the simulations.}
\label{fig:alpha_flux}
\end{figure*}

The left panel of Figure~\ref{fig:alpha_flux} illustrates these fits for two representative redshifts (z = 0.1 and 6) and two evolutionary times (t = 0.16 Myr and 0.82 Myr), demonstrating the power-law behaviour of the simulated spectra 
in the optically thin regime and the evolution of the slope with both age and redshift.
The right panel of Figure~\ref{fig:alpha_flux} shows the spectral index $\alpha$ as a function of redshift for the Rg5 simulations. 

At early evolutionary times (0.16~Myr), the spectral index exhibits only modest variation with redshift, reflecting the dominance of freshly accelerated electrons that have not yet undergone substantial radiative ageing.
In contrast, at later times (0.82~Myr), the $\alpha$ steepens significantly with increasing redshift, reaching values $\alpha \lesssim -1.2$ at $z \gtrsim 5$. 
As the system evolves, a substantial fraction of the observable emission originates from the cocoon (Figure~\ref{Radiative Intensity}), where relativistic electrons are advected away from acceleration sites and accumulate over time. 
In this region, at high redshift, the increased CMB energy density significantly enhances the IC/CMB losses and leading to more rapid spectral ageing. 
Consequently, older electron populations in the cocoon develop steeper synchrotron spectra, producing the observed redshift-dependent steepening. 
This effect becomes prominent only after sufficient evolutionary time has elapsed for cooled electrons to dominate the emission, whereas at early times 
the spectrum remains largely insensitive to redshift due to the prevalence of freshly accelerated particles. This may explain why compact and potentially younger high-redshift radio quasars do not always exhibit a significant $\alpha$-$z$ correlation observationally \citep{2021MNRAS.508.2798S}.

From an observational perspective, steep radio spectra in high-redshift sources have been reported. The parsec-scale jet PSO~J0309+27 at $z=6.1$ reveals steep-spectrum jet components with $\alpha \sim -0.84$ \citep{2020A&A...643L..12S}. Subsequent observations by \citet{Ighina2022} further revealed extended X-ray jet emission on kiloparsec scales spatially associated with extended radio emission detected in the VLASS 3\,GHz observations.

Multi-frequency VLA observations of PSO~J352.4034$-$15.3373 at $z=5.832$ also show a pronounced steep spectral profile and clear evidence for a synchrotron break, enabling estimates of the magnetic field strength and jet spectral age \citep{2025ApJ...985...34R}. 

Indeed, our simulations reproduce a similar spectral steepening behaviour, demonstrating that the CMB energy density increasing 
with redshift will enhance the IC/CMB losses and thus steepen the optically thin synchrotron spectrum. 
 These results therefore suggest that radiative losses alone can contribute significantly to the emergence of an $\alpha$–$z$ trend in the derived jet emission, 
even without invoking additional environmental effects, such as a variation in the ambient density or jet–environment interactions, 
which may further influence the jet dynamics and the particle energy evolution.

\section{Discussion}\label{sec:5}
In this section, we discuss the implications of our results, focusing on how the coupled evolution of radio and X-ray emission from relativistic jets depends on redshift.

\subsection{X-ray Enhancement and Multiwavelength Comparison}
X-ray enhancement in high-redshift radio-loud quasars reflects the increasingly rapid growth of X-ray emission relative to radio emission with redshift
\citep{Wu2013, Zhu2018, Zhu2020, Connor2021, Ighina2022, Zuo2024}. 
While the jet radio emission is dominated by synchrotron radiation and exhibits a relatively weak dependence on redshift, 
the X-ray emission is expected to increase strongly with redshift due to the growing importance of IC scattering of jet particles by CMB photons.

Our model approach enables us to quantify this behavior by jointly examining the redshift evolution of radio and X-ray luminosities resulting from our simulations,
and by comparing them with observed data \citep{Zhu2020, Zuo2024}.
For this comparison, we calculate the jet emission at a de-projected jet length of 12 kpc for the simulated blazar sources with a viewing angle of $1^\circ$. This scale is arbitrary and may not necessarily correspond to jet size in the observational samples.
We rather adopted it as a representative reference scale that enables a consistent assessment of the luminosity evolution with redshift within our simulation framework.

At this point, we do not yet incorporate observational uncertainties in the comparison, as our main objective is to evaluate qualitative trends. 
Also, explicitly, we do not account for the evolutionary stage of the quasars in the observational samples.
Therefore, our comparison is intended to highlight essential global multiwavelength trends, rather than establish a direct correspondence between individual simulated jets and specific observed sources.

\begin{figure*}
    \centering
    \includegraphics[width=0.48\linewidth]{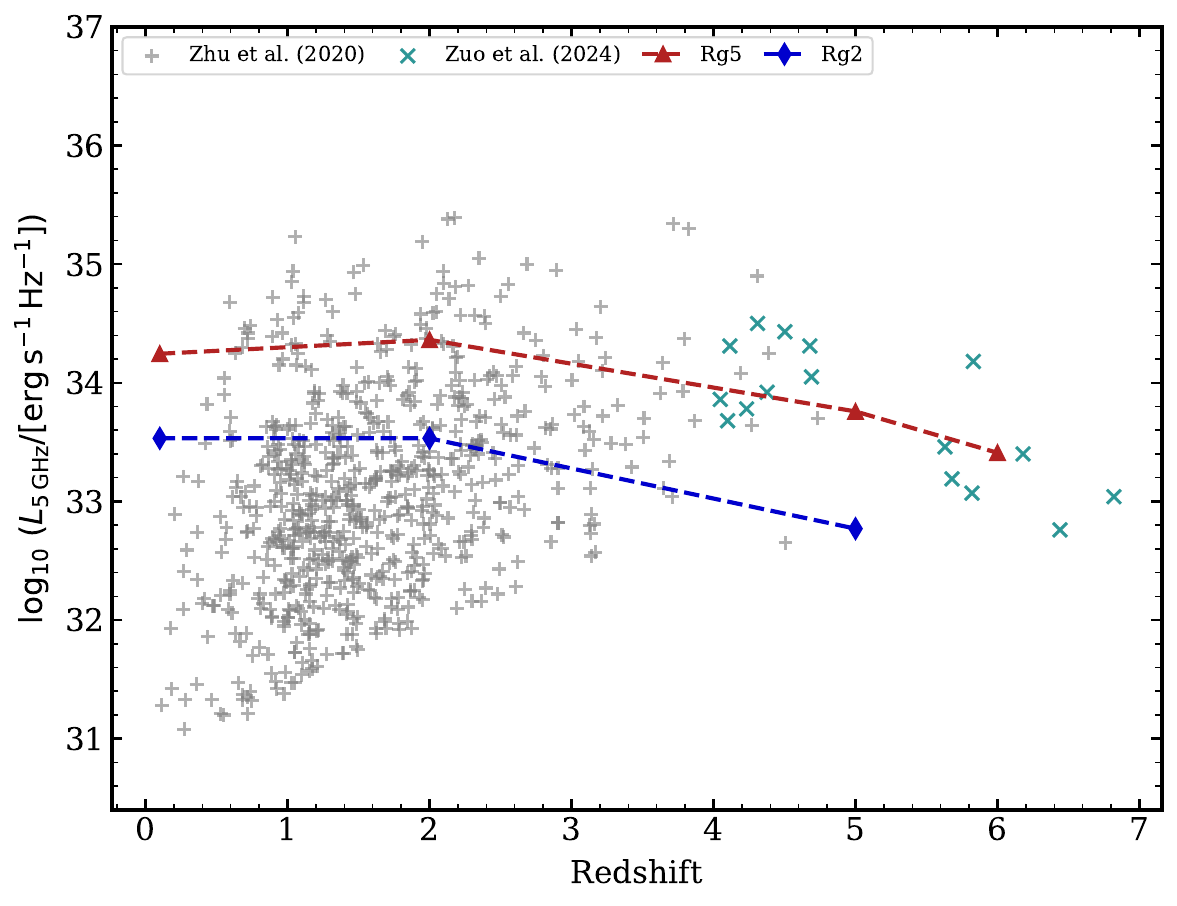}
    \includegraphics[width=0.48\linewidth]{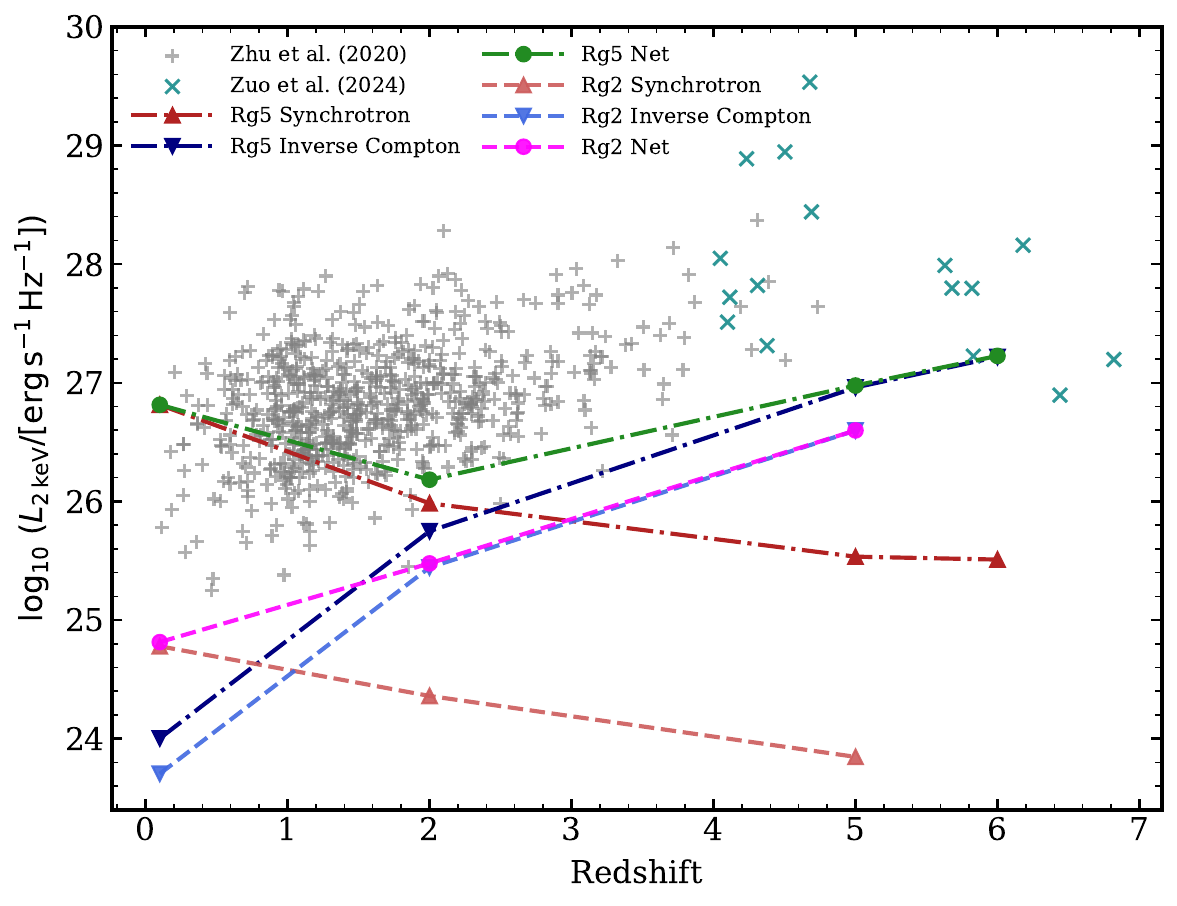}
    \caption{Comparison of simulated and observed radiative properties of jets with de-projected lengths of $\sim$12~kpc. 
    Left panel: Synchrotron radio luminosity at observed 5~GHz compared with observational data from \citet{Zhu2020} and \citet{Zuo2024}. Observational data from \citet{Zhu2020} are shown as grey plus symbols, while those from \citet{Zuo2024} are shown as teal crosses. The red and blue dashed lines represent the simulated radio luminosities for the Rg5 and Rg2 runs, respectively.
    Right panel: X-ray luminosity at observed 2~keV compared with observations from \citet{Zhu2020} and \citet{Zuo2024}. The simulated emission components are shown as synchrotron (triangles), inverse Compton (inverted triangles), and net (circles), where the total emission includes contributions from both synchrotron and IC/CMB processes. The Rg5 and Rg2 simulations are indicated by dash-dotted and dashed lines, respectively. Observational data are shown with the same symbols and colors as in the left panel. All quantities are given in the observer's frame.}
    \label{fig:Luminosity_compare}
\end{figure*}

The left panel of the Figure~\ref{fig:Luminosity_compare} shows the evolution of the simulated synchrotron radio luminosity at 5~GHz as a function of redshift for jets with a de-projected length of 12~kpc. 
The radio luminosity remains approximately constant up to $z \approx 2$, followed by a decline toward higher redshifts. 
The radio emission is produced by synchrotron radiation, which depends on the magnetic field energy density which does not increase with redshift in the same way as the CMB energy density. 
Therefore, the synchrotron emission does not experience any similar boost, and due to cosmological effects (e.g., redshift dimming), the observed radio emission decreases with redshift. 
This behavior broadly reproduces the decreasing trend observed in the radio luminosities of powerful radio-loud quasars, indicating that the simulations capture the typical radio power and its weak redshift dependence.

In contrast to the radio emission, the X-ray luminosity exhibits a much stronger dependence on redshift. 
Right panel of the Figure~\ref{fig:Luminosity_compare} presents the simulated X-ray luminosity at 2~keV as a function of redshift for jets of the same de-projected length.
At low redshift ($z \lesssim 2.5$), the X-ray emission is dominated by the synchrotron component, with luminosities consistent with those observed in most sources.
As redshift increases, the inverse Compton contribution rises steeply   
due to the rapid increase in the CMB energy density, overtaking the synchrotron component at $z \approx 2.4$.
As a result, the total X-ray luminosity increases toward high redshift, successfully reproducing the X-ray enhancement reported in observational studies.
The increase in the X-ray flux in the Rg2 simulation is broadly consistent with the expected $(1+z)^4$ scaling of the IC/CMB process after a certain redshift.

However, the total X-ray emission in our case includes contributions from both synchrotron and IC/CMB processes. 
For the Rg2 simulation, the apparent monotonic increase seen in the current plot (pink dashed curve) mainly results from the limited number of sampled redshift points. 
If we consider the Rg2 net X-ray luminosity between redshifts $z = 0.1$ and 2, we would expect a similar pattern to the Rg5 simulation, depending on the relative contributions of the different processes. With a finer sampling in redshift, the trend would likely deviate from a simple apparent monotonic increase of X-ray luminosity.

\begin{figure}
    \centering
    \includegraphics[width=1\linewidth]{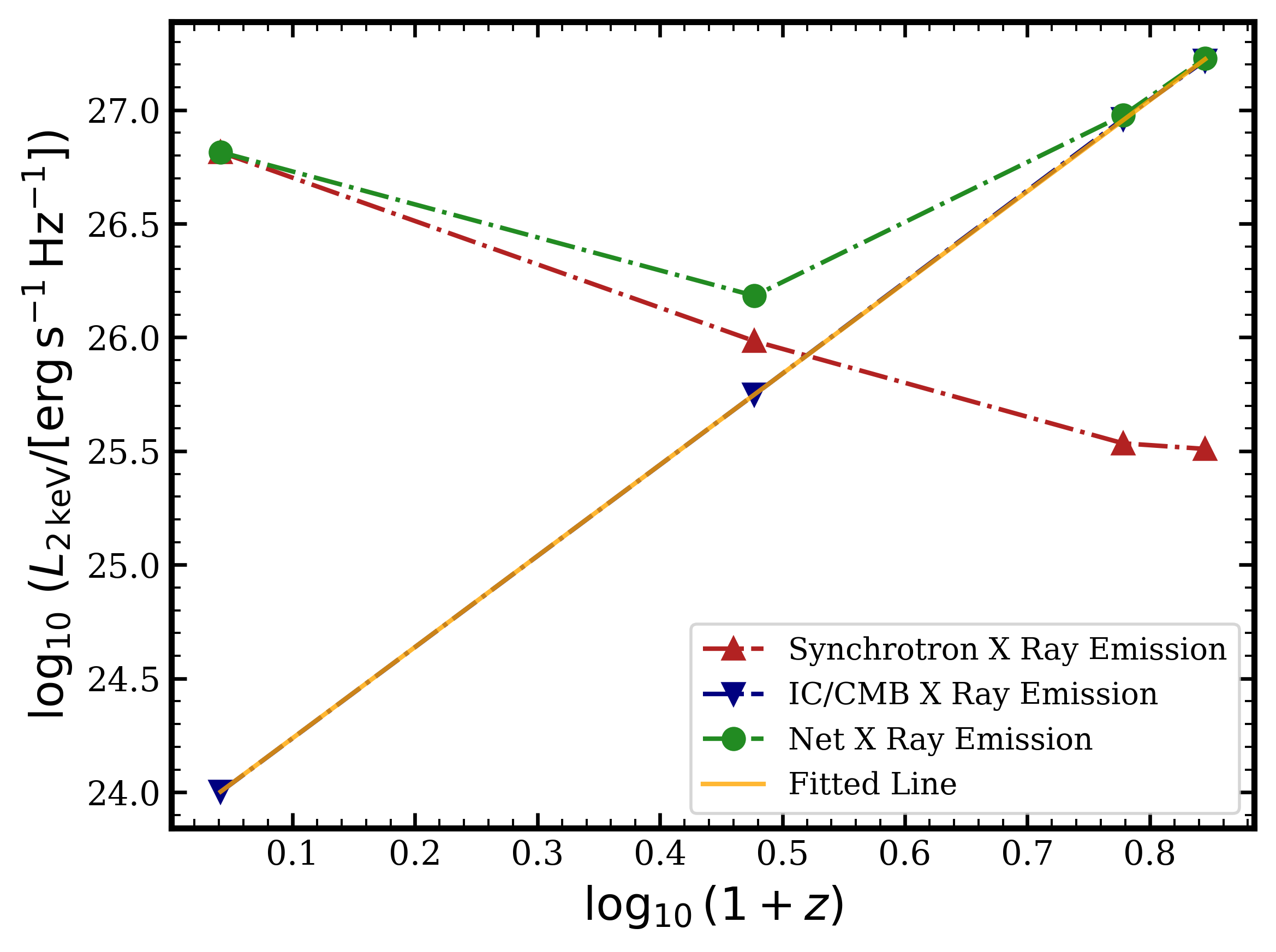}
    \caption{Evolution of the X-ray luminosity at 2~keV with redshift for the Rg5 jet simulation at an evolutionary time of 0.82~Myr. Shown are the contributions by synchrotron radiation (red), IC/CMB (blue dashed), and the net X-ray emission (green), together with a linear fit to the IC/CMB luminosity (orange), consistent with the expected $(1+z)^4$ scaling.}
    \label{Lx_fitting}
\end{figure}

To further quantify the redshift dependence of the X-ray emission, we examine how the simulated X-ray luminosity scales with $(1+z)$. Figure~\ref{Lx_fitting} shows the evolution of the X-ray luminosity at 2~keV for the Rg5 jets at a representative evolutionary time of $0.82$ Myr. At high redshift, where the inverse Compton component dominates, the total X-ray luminosity closely follows a $(1+z)^4$ dependence, as expected for IC/CMB emission. This result provides direct confirmation that the rise in X-ray luminosity with redshift is driven by the increasing CMB energy density.

\begin{figure}
    \centering
    \includegraphics[width=1\linewidth]{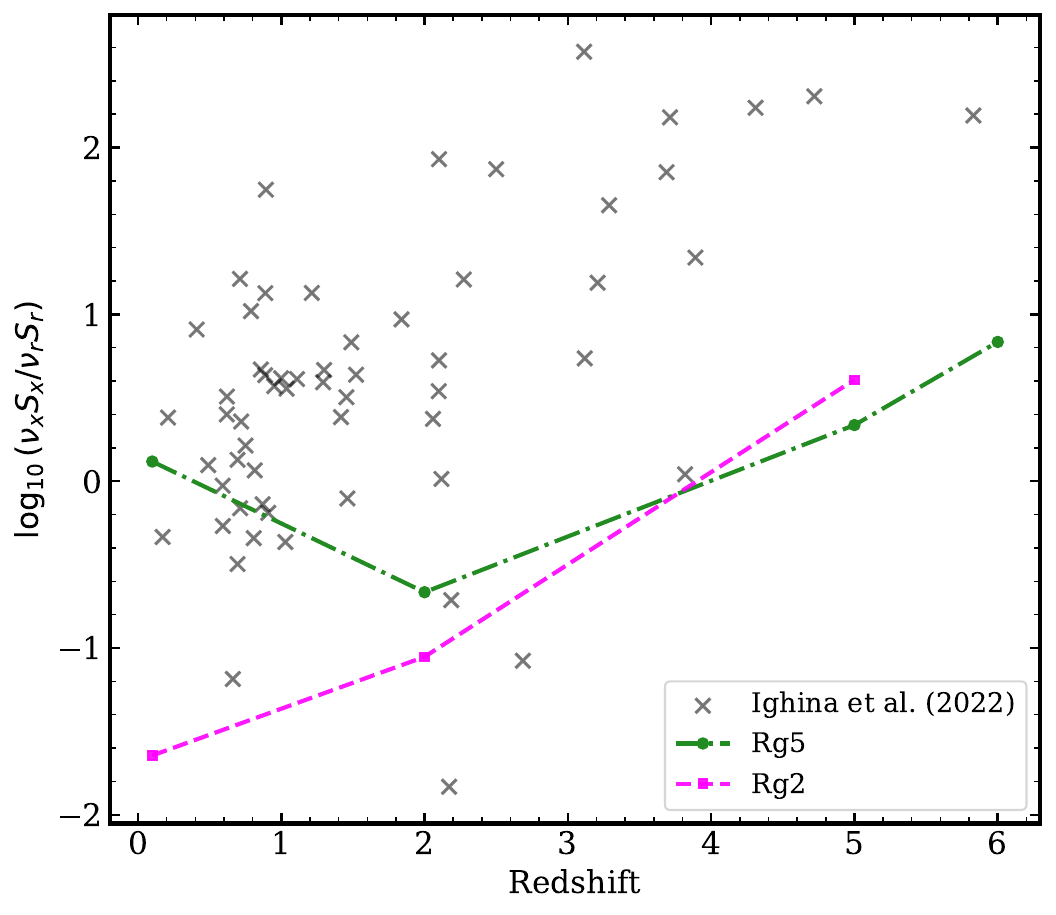}
    \caption{Evolution of the X-ray-to-radio flux ratio, $\log(\nu_X S_X / \nu_R S_R)$, evaluated at 1~keV for the X-ray band and 3~GHz for the radio band, as a function of redshift. The curves show the simulated ratios for relativistic jets with de-projected lengths of 12~kpc for the Rg5 (green, dash-dotted) and Rg2 (magenta, dashed) models. Observed data points from \citet{Ighina2022} are overplotted for comparison. The systematic increase of the flux ratio with redshift reflects the growing importance of IC/CMB emission at high redshift.}
    \label{fig:XR_redshift}
\end{figure}

The different redshift evolution of the radio and X-ray luminosities naturally leads to an increase in the X-ray-to-radio flux ratio at high redshift. 
To quantify this enhancement, we compute the X-ray-to-radio flux ratio, $\nu_X S_X / \nu_R S_R$, evaluated at 1~keV for the X-ray band and 3~GHz for the 
radio band, for our simulated jets and compare the results with the observational sample of \citet{Ighina2022}. 
Figure~\ref{fig:XR_redshift} shows the evolution of the flux ratio for relativistic jets with de-projected length of 12~kpc for the Rg5 and Rg2 models. 
In both cases, $\nu_X S_X / \nu_R S_R$ increases systematically with redshift, reflecting the growing dominance of IC/CMB emission over synchrotron processes.

We find that the stronger redshift dependence of the X-ray-to-radio ratio in the Rg2 simulations as compared to the Rg5 case 
arises from differences in jet evolution and particle energetics.
In particular, because the Rg2 jets propagate more slowly, they require more time to reach the same spatial extent, leading to enhanced cumulative radiative 
and adiabatic cooling of the electron population. 
Although particle acceleration and continuous injection persist throughout the jet evolution, these processes are spatially localized, whereas cooling acts 
continuously on the entire electron population, as it is advected into the cocoon and backflow regions. 
As a result, the Rg2 jet simulations contain a larger number of electrons across the full energy distribution, including both low- and high-energy populations. 

However, despite this larger particle content, the synchrotron radio emission is less efficient due to the lower magnetic energy density 
(Table~\ref{tab:initial_param}), leading to a more rapid decline in the radio luminosity. 
At the same time, the X-ray emission remains strong due to IC/CMB processes. 
Consequently, the combination of suppressed radio emission and sustained X-ray production naturally produces the steeper evolution of the X-ray-to-radio 
flux ratio in the Rg2 simulations.

\subsection{Low-Frequency Turnover in Radio Sources}
\begin{figure}
    \centering
    \includegraphics[width=1\linewidth]{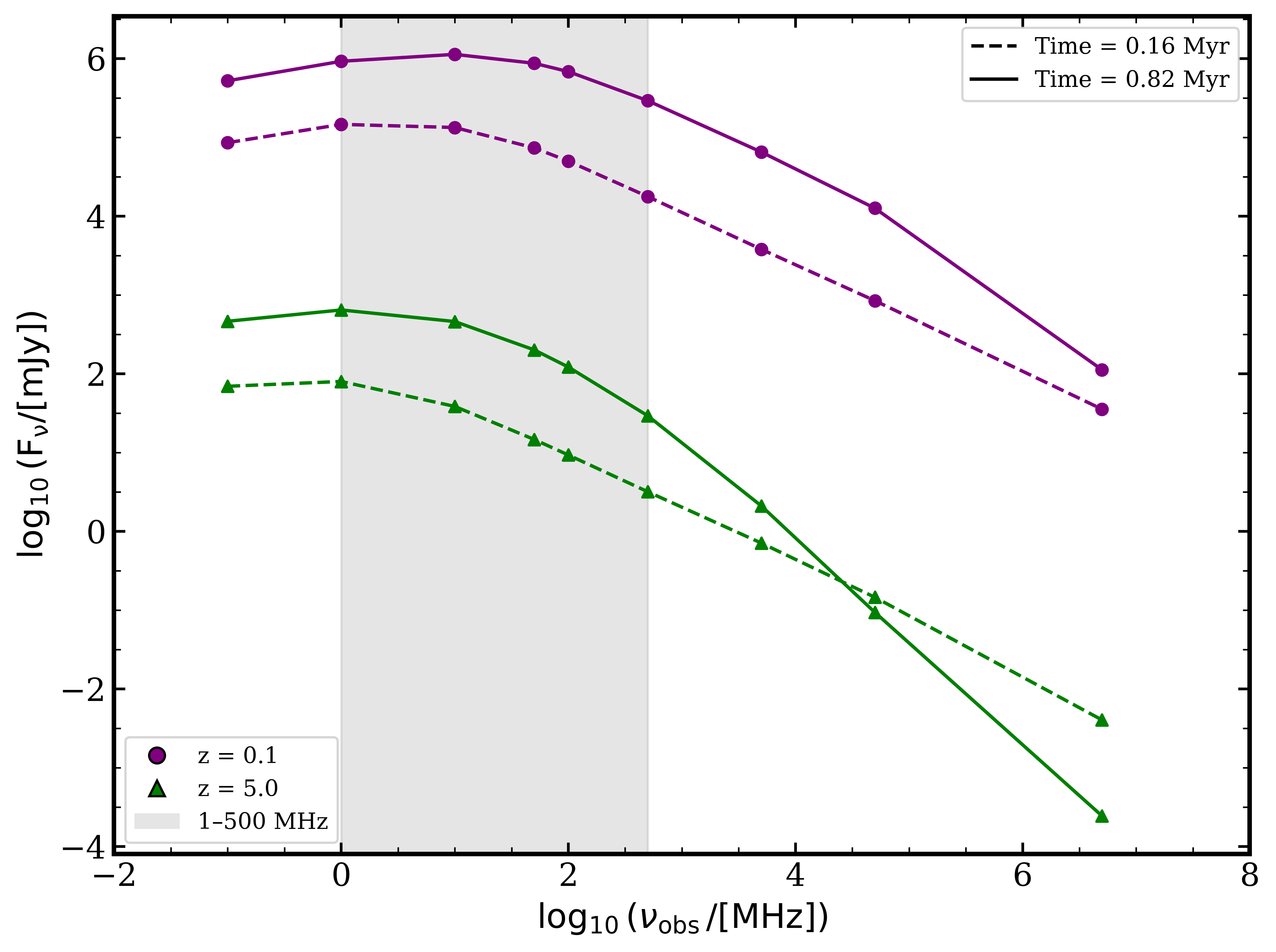}
    \caption{Simulated radio flux density as a function of observing frequency for the Rg5 jet simulations at two evolutionary stages, $t = 0.16$~Myr (dashed lines) and $t = 0.82$~Myr (solid lines). Symbols indicate different source redshifts: $z = 0.1$ (purple circles), and $z = 5.0$ (green triangles). 
    The gray shaded region marks the low-frequency band between 1 and 500~MHz.}
    \label{Flux vs freq}
\end{figure}

Figure~\ref{Flux vs freq} shows the radio spectra obtained from our simulations at redshift 0.1 and 5 for Rg5 simulations. 
In our runs, the turnover frequency is primarily determined by the low end of the electron energy distribution, particularly the choice of the minimum Lorentz factor, $\gamma_{\min}$. 
We note that particles are injected with a minimum Lorentz factor of $\gamma_\mathrm{min} = 10^2$. 
With time, this minimum Lorentz factor can change due to both particle acceleration and radiative cooling processes. 
As particles cool, some of them may reach $\gamma_\mathrm{min} < 10^2$, and such particles remain in the computational domain. 
These lower-energy particles mainly contribute to the lower-energy part of the emission spectrum and therefore regulate the low-frequency flux. 
Smaller values of $\gamma_{\min}$ reduce the number of electrons radiating at higher radio frequencies and therefore shift the turnover to lower frequencies. 
Because $\gamma_{\min}$ evolves through particle acceleration and subsequent radiative cooling, the resulting turnover frequency naturally 
reflects the dynamical state of the jet.

It is important to note that the present simulations do not include synchrotron self-absorption (SSA) or external free-free absorption (FFA), both of which may significantly influence the radio spectra of compact sources. 
Observational modeling of radio-loud quasars at $z>5$ has shown that FFA in an external inhomogeneous medium can accurately reproduce the observed peaked spectra in most cases, 
which may indicate an FFA origin for the radio spectral turnovers in these sources.
Many observed high-redshift radio-loud quasars also exhibit spectral peaks in the GHz frequency regime \citep{2015MNRAS.450.1477C, 2022A&A...659A.159S}, which are likely influenced by synchrotron self-absorption, external free-free absorption, 
compact source structure, and dense ambient environments. In addition, multiple spectral components and source variability may further influence the observed spectral structure \citep{2025NatAs...9..293B, Belladitta2026}. 
In contrast, the turnovers in our simulations arise solely from the evolution of the lower bound of the electron energy (i.e., $\gamma_\mathrm{min}$). This is evolved via consistently modeling effects of radiative losses and shock acceleration processes.

Our results therefore demonstrate that intrinsic particle-distribution effects can alter the spectral curvature in low radio frequency regime, even in the absence of absorption processes. 
However, the specific details of the optically thick spectral slope and the physical origin of the turnover due to SSA and FFA cannot be directly constrained from this study.
Nevertheless, the turnover frequencies as resulting from our simulations remain below $\sim 500$~MHz across the explored redshift range. 
This is consistent with the low-frequency spectral curvature observed in compact radio sources over a broad redshift interval 
\citep{2015MNRAS.450.1477C, 2017ApJ...836..174C}. Furthermore, for $z \gtrsim 6$, sources with rest-frame turnovers below $\sim 500$~MHz would appear at observed frequencies below $\sim 70$~MHz. Such low frequencies are difficult to probe with current facilities and will likely require next-generation instruments such as SKA-Low.
 
In a subsequent study, we shall aim to better constrain the origin of spectral turnovers by including the effects of SSA and FFA, together with the potential impact of variations in the ambient medium. Incorporating these absorption processes within a full radiative transfer framework will be essential for distinguishing between intrinsic particle-evolution-driven turnovers and absorption-dominated scenarios in high-redshift radio quasars.

\section{Summary}
We have studied the evolution of radio and X-ray emission from relativistic AGN jets that are launched at different cosmic time, with particular emphasis on the role of the CMB radiation. In particular, we investigate how IC scattering of CMB photons by relativistic electrons influences the radiative output of jets and regulates the relative strengths of radio and X-ray emission as a function of redshift.

For this work, we have extended, tested, and updated the existing hybrid Eulerian–Lagrangian framework of the PLUTO code by developing and incorporating a new radiation module that considers inverse Compton scattering of the CMB background, and thus allows to examine the effects of a varying CMB radiation density with redshift. The framework enables a self-consistent treatment of the coupled dynamical and radiative evolution of the AGN jet, providing us with synthetic multiwavelength emission maps and spectra, while preserving the intrinsic jet dynamics across redshift.

Our study is based on three-dimensional relativistic magnetohydrodynamic (RMHD) simulations of fully relativistic, rotating, and magnetized jets performed using the PLUTO code.
We investigate jet bulk Lorentz factors $\Gamma_{j,c}=5$ and $\Gamma_{j,c}=2$, corresponding to jet powers of $1.64 \times 10^{44}\,\mathrm{erg \, s^{-1}}$ and $1.7 \times 10^{43} \, \mathrm{erg \, s^{-1}}$, respectively,
and compare redshifts of $z = 0.1,\, 2, \, 5$ and $6$. 
The synthetic emission is calculated for a viewing angle of $1^\circ$, corresponding to a blazar-like line of sight.
Essentially, we consider kpc-scale jets, performing our simulations in a computational domain of $6 \,\mathrm{kpc} \times 6 \,\mathrm{kpc} \times 16 \,\mathrm{kpc}$.

By keeping the jet dynamics fixed for a given set of injection parameters and varying only the CMB photon energy density according to its cosmological scaling, $(1+z)^4$, we have been able to disentangle the impact of redshift-dependent radiative processes from dynamical effects. 
This approach allows us to systematically study how the radiative signatures evolve with redshift, as well as their dependence on jet age and spatial scale along the jet.

We summarize our results as follows.

\begin{enumerate}

    \item We show that the relative contributions of synchrotron and IC/CMB emission to the X-ray band depend sensitively on both the jet dynamical evolution and the source redshift.
    
   \item The X-ray emission mechanism evolves systematically with jet age: young kpc-scale jets remain synchrotron dominated because freshly injected and shock-accelerated electrons near the jet spine retain sufficiently high energies, whereas radiative cooling progressively suppresses this component in older jets, leading to a transition toward IC/CMB-dominated X-ray emission.

    \item In evolved and extended jets, a significant fraction of the IC/CMB emission originates from the cocoon and backflow regions, where relativistic electrons accumulate over time and undergo comparatively weak re-acceleration.
    
    \item The radio emission remains synchrotron dominated throughout the evolution and declines more gradually than the X-ray synchrotron component. The simulations naturally reproduce both the observed decrease in radio luminosity and the progressive steepening of the spectral index with redshift, even for intrinsically identical jets, without requiring changes in the intrinsic jet launching conditions or ambient density evolution.
    
    \item Jets with lower Lorentz factor exhibit a stronger X-ray enhancement relative to the radio emission compared to the faster jets, driven by their slower propagation, enhanced cumulative cooling, and lower magnetic energy density. This highlights the important role of jet dynamics in shaping the observed X-ray-to-radio evolution.
    
    \item The simulations naturally produce spectral turnovers below $\sim 500$~MHz through the evolution of the low-energy electron distribution, even in this simplified case with the absence of synchrotron self-absorption or external free-free absorption. 
    
\end{enumerate}

Overall, this work demonstrates that the interplay between the jet evolution and the cosmological radiative environments plays a key role in shaping the broadband emission properties of relativistic jets and must be taken into account when interpreting radio and X-ray observations, particularly at high redshift.\\

\begin{acknowledgments}
We thank the anonymous referee for valuable suggestions that helped improve the clarity and quality of this paper.
A.S. acknowledges funding from the Council of Scientific and Industrial Research (CSIR) via a CSIR-JRF fellowship, under the grant 09/1022(17697)/2024-EMR-I. 
B.V. would like to acknowledge the support from the Max Planck Partner group Award established at IIT Indore.
B.V., H.B. and C.F. acknowledge support of the Deutsche Forschungsgemeinschaft (DFG, German Research Foundation), project number 443220636, via the Research Unit FOR\,5195. 
D.V.L. acknowledge the support of the Department of Atomic Energy, Government of India, under project no. 12-R\&D-TFR-5.02-0700.
All the simulations have been performed using the MPCDF computing cluster Vera of the Max Planck Society and the facilities provided at IIT Indore.
\end{acknowledgments}

\appendix

\section{Injected Jet Profile}
\label{Appendix 1}
For the setup of our injection nozzle, we implemented the equilibrium configuration as outlined by \cite{Bodo2019}. 
In this setup, the jet propagates along the $z$-axis with no radial velocity component, such that the velocity and magnetic field profiles take the form $(0, v_\phi(r), v_z(r))$ and $(0, B_\phi(r), B_z(r))$, respectively. The magnetic field structure is characterized by the pitch parameter, 
\begin{align}
    P = r \frac{B_z(r)}{B_\phi(r)} ,
\end{align}
which quantifies the relative contribution of the poloidal and toroidal components. \\
Under these assumptions, only non trivial equation that exists after this simplification is the radial component of the momentum equation, and can be written as
\begin{align}\label{eq:momentum_balance}
    \rho \gamma^2 v_\phi^2 = \frac{1}{2r} \frac{d}{dr}\!\left(r^2 H^2\right) 
    + \frac{r^2}{2} \frac{d B_z^2}{dr} ,
\end{align}
where $H^2 = B_\phi^2 - E_r^2$ and $E_r = v_z B_\phi - v_\phi B_z$ is the radial electric field. \\
To solve for the jet's structure, we first define the radial profiles for proper density and velocity, which are then used with the radial momentum equation to find the remaining variables. These initial profiles are constructed to vary within the jet's radius.
The assumed profile for the $z$-component of the velocity, via bulk Lorentz factor ($\Gamma_z$) of the jet, is
\begin{align}
    \Gamma_z(r) = 1 + \frac{\Gamma_c - 1}{\cosh^6(r/r_j)}
\end{align}
where $\Gamma_c$ is the bulk Lorentz factor on the axis ($r=0$).
The density profile is
\begin{align}
    \rho_j(r) = \eta + \frac{1 - \eta}{\cosh^8(r/r_j)}
\end{align}
where $\eta$ is the density contrast between ambient and jet.
They assign a profile to H as,
\begin{align}\label{eq:H2}
    &H^2 = \frac{H_c^2}{ r^2 }\left[ 1 - \exp \left(-\frac{r^4}{a^4} \right) \right]
\end{align}
and the azimuthal velocity $v_{\phi}$ is determined by the jet's rotation, characterized by the angular velocity at the axis, $\Omega_c$. 
The resulting profile is given by
\begin{align}\label{eq:vphi}
     &v_\phi^2 = \frac{r^2 \Gamma_c^2 \Omega_c^2 }{\Gamma_z^2}\left[1 + r^2 \Gamma_c^2 \Omega_c^2 \exp \left(-\frac{r^4}{a^4} \right) \right]^{-1}\exp \left(-\frac{r^4}{a^4} \right)
\end{align}
where $a=0.6r_j$ is a characteristic radius concentrating the current density inside the jet. 
This equation ensures that $v_{\phi}$ is always less than unity for any $\Omega_c$.

From the equation \ref{eq:momentum_balance}, \ref{eq:H2} and \ref{eq:vphi}, we get the axial magnetic field, $B_z$ as
\begin{align}
    B_z^2 = B_{zc}^2 - (1 - \alpha)\,\frac{H_c^2 \sqrt{\pi}}{a^2}\,\mathrm{erf}\!\left(\frac{r^2}{a^2}\right)
\end{align}
where erf is the error function and the parameter, 
\begin{align}
    \alpha = \frac{\rho \Gamma_c^2 \Omega_c^2 a^4}{2 H_c^2}
\end{align}
which measures the strength of the centrifugal force relative to the magnetic pressure gradient. 
For this study, we consider a rotating jet where $\alpha=1$, resulting in a constant axial magnetic field
\begin{align}
    B_z(r) = B_{zc}.
\end{align}
The toroidal magnetic field, $B_{\phi}$, is then found by solving for the magnetic field components. We choose the negative branch of the solution to ensure the toroidal field and azimuthal velocity are oppositely directed, consistent with acceleration models \citep{Blandford1982}:
\begin{align}
     B_\phi &= \frac{-v_\phi v_z B_z \pm\sqrt{v_\phi^2 B_z^2 + H^2(1 - v_z^2)}}{1 - v_z^2}
     \label{b_phi profile}
\end{align}

As mentioned above, the pitch parameter is used to characterize the configuration of the magnetic field, and it is defined near the jet axis ($r\to0$) as
\begin{align}
\label{1}
     &P_c \equiv \left|\frac{r B_z}{B_\phi}\right|_{r=0}.
\end{align}
Another parameter that is used to characterize the magnetic field configuration is $M_a^2$ (alfvenic mach number), defined as the ratio of the matter-energy density to the magnetic energy density, 
\begin{align}
\label{2}
      &M_a^2 \equiv \frac{\rho_j \Gamma^2_c}{\langle B^2\rangle}
\end{align}
where $\langle B^2 \rangle$ denotes the average magnetic field in the radial direction
 \begin{align}
 \label{3}
     &\langle B^2 \rangle =\frac{\int_{0}^{r_j} (B_z^2 + B_\phi^2) r dr}{\int_{0}^{r_j} r dr}, 
 \end{align}
Finally, the constants $B_{zc}$ and $H_c$ are determined by simultaneously solving the system of equations \eqref{1}-\eqref{3}, which in the limit $r\to 0$ yields the following relationship between the parameters,
\begin{align}
    a^4B_{zc}^2 = \frac{H_c^2 P_c^2}{1 - (P_c \Omega_c + v_{zc})^2}.
\end{align}
For a solution to be physically meaningful, the magnetic energy terms $B_z^2$ and $B_{\phi}^2$ must remain positive throughout the domain. 
This requirement places constraints on the allowable combinations of the free parameters ($\Omega_c$, $\Gamma_c$, $P_c$, $M_a$).

\begin{figure}
    \centering
    \includegraphics[width=1\linewidth]{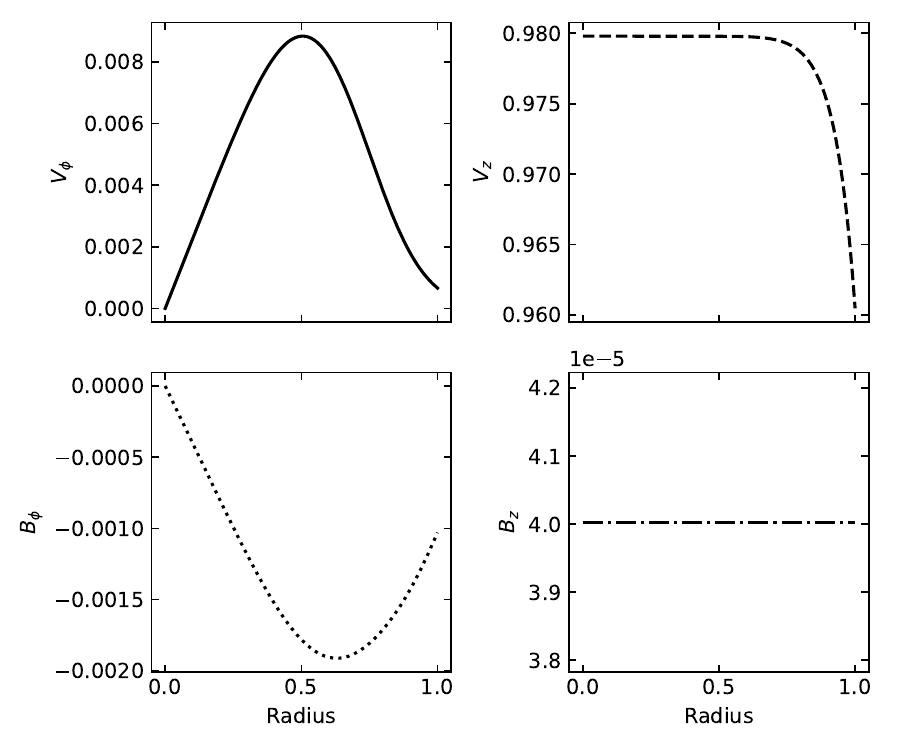}
    \caption{This figure displays the radial profiles for various physical quantities within the jet for Rg5 simulation (in code units).}
    \label{fig:radial_profiles}
\end{figure}

The complete structure of the magnetized, rotating relativistic jet is therefore determined by the choice of these governing parameters. The specific values employed in this study are summarized in Table~\ref{tab:initial_param}, while the corresponding radial profiles of the ($v_z$), azimuthal velocity ($v_\phi$), and magnetic field components ($B_z$, $B_\phi$) are shown in Figure~\ref{fig:radial_profiles} for Rg5 simulation.

\section{One Zone Model}
\label{Appendix 2}

To validate the modifications implemented in the PLUTO particle module, we performed a 3D one-zone simulation and generated the corresponding synthetic spectral energy distribution (SED). 
The simulation parameters used are summarized in Table \ref{tab:one_zone_table}.
For consistency, we also computed an independent SED using the open-source python package AGNpy \citep{Nigro2022}, employing identical physical parameters.

\begin{table}
\centering
\caption{Input Parameters Used in the One Zone Model Simulation}
\begin{tabular}{lc}
\hline\hline
\textbf{Parameter} & \textbf{Value} \\
\hline
Redshift ($z$) & $5.0$ \\
Magnetic field strength ($B_{\mathrm{blob}}$) & $100~\mu\mathrm{G}$ \\
Bulk Lorentz factor ($\Gamma_{\mathrm{bulk}}$) & $10.0$ \\
Blob radius ($R_b$) & $9.257 \times 10^{24}~\mathrm{cm}$ \\
Viewing angle ($\Theta$) & $1.0^{\circ}$ \\
Spectral index ($p$) & $6.0$ \\
Minimum Lorentz factor ($\gamma_{\min}$) & $10^{2}$ \\
Maximum Lorentz factor ($\gamma_{\max}$) & $10^{8}$ \\
Number of energy bins ($\gamma_{\mathrm{bins}})$ & $4096$ \\
Electron Number Density ($N_e$) & $4.95 \times 10^{-5}~\mathrm{cm^{-3}}$ \\
Time ($t_{\mathrm{end}}$) & $6.48 \times 10^7~\mathrm{s}$ \\
\hline
\end{tabular}
\label{tab:one_zone_table}
\end{table}

\begin{figure}
    \centering
    \includegraphics[width=1.0\linewidth]{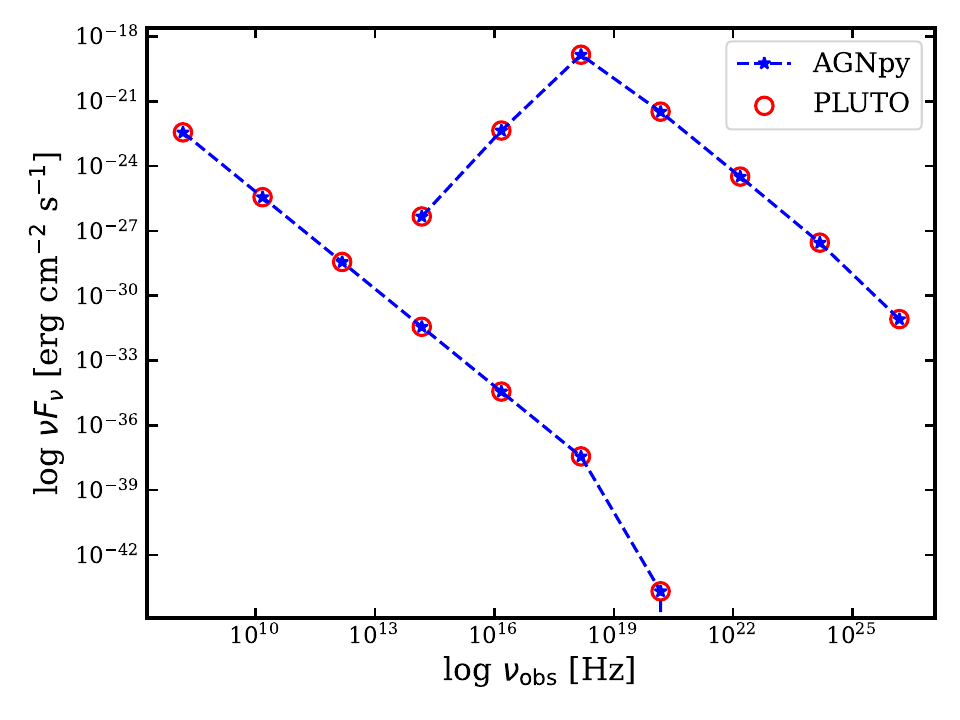}
    \caption{Comparison of the spectral energy distributions (SEDs) computed using AGNpy (lines) and PLUTO (open circles) for synchrotron and IC/CMB emission components.}
    \label{SED Comparison}
\end{figure}

As shown in Figure~\ref{SED Comparison}, the SED obtained from PLUTO is in excellent agreement with the AGNpy results across the entire frequency range, with the relative percentage error less than 2\%. The minor discrepancies likely arise from differences in the numerical integration schemes adopted by the two approaches. This consistency demonstrates the robustness and reliability of the modified particle radiation module.

\bibliography{paper_references}{}
\bibliographystyle{aasjournalv7}

\end{document}